# Droplet encapsulating bubble: Investigation of droplet spreading dynamics and bubble encapsulation time

## A PREPRINT

**Adel Ebadi[1], Raha Kalantarpour[1,2], Fariborz Ataei[1], Hesam Ami Ahmadi[1], S.M. Hosseinalipour[1*]**

[1] School of Mechanical Engineering, Iran University of Science and Technology, Tehran, Iran.

[2] Department of Mechanical Engineering, University of California, Riverside, Riverside, CA 92521, USA

*Corresponding author: Postal Code: 13114-16846, Tehran, Iran

Email: adel_ebadi@alumni.iust.ac.ir

Email: alipour@iust.ac.ir

(Dated: 25 April 2025)

Ternary interactions between hetero-fluid particles, particularly the dynamics of droplets spreading over curved fluid interfaces remain insufficiently understood compared to the two-phase coalescence. In this study, we combine lattice Boltzmann simulations, high-speed imaging, and theoretical scaling to investigate the collision and encapsulation of an air bubble by a rising oil droplet in an immiscible medium. We systematically vary fluid properties, droplet-to-bubble size ratios, and collision configurations to quantify their impact on encapsulation time and flow evolution. The process unfolds in four stages: collision/film drainage, encapsulation, reshaping, and compound rising. Results indicate that encapsulation time increases exponentially with viscosity and is strongly modulated by the spreading coefficient ($S_o$), which governs the imbalance of interfacial tensions. Higher $S_o$ values enhance capillary-driven spreading and reduce engulfment duration, while lower values yield coupled deformation-reshaping behavior, introducing oscillations in bubble velocity and shape evolution. For low viscosity drops ($Oh<0.1$), the neck growth rate follows the well-known power-low relation with an exponent of 0.44-0.5, dependent on the size ratio. The transition between the spherical and deformed regimes is identified. Our theoretical analysis reveals that in low Bond numbers ($Bo<0.11$), spreading speed scales with viscous-capillary velocity, while in the deformed regime ($0.11<Bo<2.2$), encapsulation time follows a capillary-gravitational timescale. Interestingly, smaller droplets expedite encapsulation in equal-sized collisions but delay it in size-mismatched pairs, despite a faster initial neck growth. These findings provide new mechanistic insight into three-phase fluid interactions and offer guidance for optimizing encapsulation processes in applications such as emulsification, gas flotation, and interfacial microfluidics.

## 1. Introduction

Three-phase/component flows involving bubbles and droplets, are ubiquitous in nature and hold pivotal roles across numerous environmental (Dudek & Øye 2018; Ji *et al.* 2021c, 2022; Maës *et al.* 2025; Piccioli *et al.* 2024; Yan *et al.* 2020), food (Duangsuwan *et al.* 2009; Won *et al.* 2014), pharmaceutical and biological (Ji *et al.* 2021b; Planchette *et al.* 2010; Potyka & Schulte 2024; Wang *et al.* 2014), microfluidics (Bariki & Movahedirad 2024; Chen *et al.* 2019b;

Fu *et al.* 2016; Geng *et al.* 2020; Park & Anderson 2012a), ultrasonic emulsification (Li *et al.* 2024; Ren *et al.* 2023), and chemical processes (Shen *et al.* 2018; Zhang *et al.* 2017, 2022, 2016). A key application motivating this study is gas flotation, a technique used in treating produced water by enhancing the rise velocity of fine oil droplets via attachment to gas bubbles (Saththasivam *et al.* 2016; Wang *et al.* 2024a). This approach highlights the critical importance of small-scale interactions, such as droplet growth via coalescence and successful bubble-droplet attachment in achieving effective separation of the dispersed oil droplets (Wang *et al.* 2014). These phenomena are particularly critical in offshore facilities, where space and time constraints drive the development of compact flotation units (CFUs). These units must ensure rapid and efficient removal of dispersed droplets while minimizing equipment size (Das & Jäschke 2019; Ding *et al.* 2023). Efficient separation critically depends on stable bubble-droplet attachment, best achieved when droplets fully encapsulate bubbles, forming oil films that resist detachment due to turbulence or collisions. The bubble-droplet attachment process is governed by interfacial phenomena. In ternary systems, the final morphology of the compound is determined by the disequilibrium of surface tensions between the three fluids, known as the spreading coefficient, $S$ (Binyaminov *et al.* 2021; Cuttle *et al.* 2021; Torza & Mason 1970). Taking $\sigma_{wg}$, $\sigma_{wo}$, $\sigma_{og}$ to represent surface tension between water-gas, oil-water, and oil-gas interfaces, respectively, the spreading coefficient of the oil droplet can be written as $S_o = \sigma_{wg} - (\sigma_{og} + \sigma_{wo})$. The full bubble encapsulation occurs when $S_o > 0$, and $S_w$, $S_g < 0$, otherwise the interaction results in partial-engulfment, also known as Janus ($S_o < 0$, $S_w < 0$, $S_g < 0$) or nun-engulfing ($S_o < 0$, $S_w > 0$, $S_g < 0$) morphologies.

Similar to homo-coalescence in two-phase systems, hetero bubble-droplet interactions involve distinct stages that are associated with different timescales (Tabor *et al.* 2011). The contact time ($t_{con}$) is typically defined as the period during which flow conditions allow fluid particles to remain in close proximity or apparent contact (Federle & Zenit 2024). The drainage ($t_{dra}$) or induction time (Eftekhardadkhah & Øye 2013) is measured from the moment of initial contact (the formation of a thin film between two particles) to the film rupture and the onset of encapsulation. The time required for the drop to spread over and totally encapsulate the bubble is referred to as the encapsulation time ($t_{enc}$), analogous to the coalescence time in the homo case. For encapsulation or coalescence to occur, the sum of the drainage and encapsulation times must be shorter than the contact time. In practice, $t_{con}$ is very short due to the dynamic flow conditions, making it critical to minimize $t_{dra}$ and $t_{enc}$. At the microscale, the result of the interaction in terms of successful film drainage is determined by the interplay between capillary and intermolecular forces, including electrostatic double-layer (EDL) and van der Waals (vdW), known as DLVO theory. The non-DLVO forces such as hydrophobic forces have been argued to play a part in the force balance to either assist the drainage or hinder it (Ami Ahmadi *et al.* 2023; Tabor *et al.* 2011). Experimental observations have revealed the effect of salt or physicochemical properties on the drainage dynamics and attachment or flotation efficiency in horizontally or vertically aligned quasi-static interactions (Ami Ahmadi *et al.* 2023; Chakibi *et al.* 2018; Eftekhardadkhah & Øye 2013; Yan *et al.* 2020). In this context, very recently, it has been also suggested that coating a condensate film layer over a bubble's surface can improve attachment time and efficiency (Lü *et al.* 2024; Wang *et al.* 2024a). On the other hand, microfluidic-type setups have been employed to mimic the dynamic interactions between the gas bubble and oil droplets to probe the role of thermophysical or physicochemical properties in the enhancement of attachment efficiency (Dudek & Øye 2018; Piccioli *et al.* 2024; Zhang *et al.* 2016). In a recent study, using a free-floating experimental setup (Wang *et al.* 2024b) it was found that droplet-to-bubble size ratios bigger than 0.75 lead to the formation of a more stable oil layer, while for smaller size ratios the attachment is partial and unstable.

The formation of the oil-coated bubbles has been investigated using experimental observations and theoretical modelling (Ji *et al.* 2021a, 2021c), indicating that the bubble encapsulation considerably affects the bubble dynamics in comparison to a single bubble rise. These studies also presented a theoretical model derived from the force balance to predict the size of the compound and its rise velocity. The hydrodynamics of rising oil-encapsulated bubbles have been the subject of numerous experiments, often accompanied by theoretical or numerical analysis (Behrens 2020; Duangsuwan *et al.* 2009; Ji *et al.* 2022; Karp *et al.* 2023; Wang *et al.* 2018), revealing that the oil-coated bubbles exhibit a rounder shape and straighter trajectory with slower ascend than bare bubble (Wang *et al.* 2018). Similarly, for the bigger droplet-to-bubble size ratios reduced drag, damped oscillations, and comparatively linear path were reported (Ji *et al.* 2022), while at lower oil fractions a zigzagging trajectory with a steady oscillation pattern was observed.

With the development and maturing of the three-phase/component fluid flow models, particularly in the LBM context, numerical simulations also have advanced understanding of ternary interactions (Boyer & Lapuerta 2006; Fu *et al.* 2016; Kalantarpour *et al.* 2020; Kim 2012; Liang *et al.* 2016; Park & Anderson 2012b; Shi *et al.* 2016; Wöhrwag *et al.* 2018; Xie *et al.* 2018; Yu *et al.* 2019a, 2019b; Zheng & Zheng 2019). In our previous study (Kalantarpour *et al.* 2020), we demonstrated that the off-center impacts leads to increased interaction and compound rising time compared to the head-on collisions. Interaction between a trailing bubble with a rising droplet has been examined as a test case in two wetting and non-wetting states using coupled NS and Phase-field approach by Shen and Li (2023). Ghorbanpour-Arani et al. (2020) considered the interaction between a rising bubble and stationary droplet with the same density as the carrier phase and reported a decrease in the drop cleanup time by reducing the drop-to-bubble density ratio, increasing bubble-to-drop diameter ratio, and increasing the viscosity ratio. In a similar setting, recently Zhao & Lee (2023) employed an imaginary buoyancy force on the rising bubble to investigate its interaction with stationary droplets using LBM simulations. These authors found that compared to the partially engulfed (Janus) droplet, double emulsion configuration leads to a higher rising velocity and more stable aggregate. A major issue with the two latter studies in (Ghorbanpour-Arani *et al.* 2020; Zhao & Lee 2023) is that the bubble phase encapsulates the droplet in their simulations, which is impractical within gas flotation application and in practice it's the other way around.

Compared to the two-phase coalescence, three-phase interactions between hetero-fluid particles have not received as much attention and the problem is far from being fully understood. In particular, the dynamics of droplets spreading over curved fluid interfaces, such as bubbles, and the associated encapsulation processes have not been thoroughly explored. To address this gap, we employ the phase-field multiple-relaxation-time lattice Boltzmann method (PF MRT-LBM) to investigate the spreading dynamics of a droplet over a bubble, both rising in an aqueous solution. The primary objective is to calculate and quantify bubble encapsulation times under various physical and kinematic conditions. A comprehensive parametric study is conducted, varying viscosity, interfacial tension, density, gravitational acceleration, collision parameters, and droplet/bubble sizes. Simulations cover both the spherical regime (typically $D_{d,b} < 1.6$ mm), where gravitational effects are negligible, and the slightly deformed regime (1.6 mm<$D_{d,b}$<2.5 mm), where gravity and the Bond number play a role. Our focus is particularly on the low-Bond-number domain ($Bo$<1), where droplet and bubble sizes fall below the capillary length, making the findings particularly relevant to practical applications such as flotation. Lattice units are converted into physical values representative of real-world conditions, and the numerical results are validated through experimental observations, which also provide further insight into encapsulation dynamics. Finally, we perform a theoretical modeling based on scale analysis to unify the results. This analysis yields theoretical relationships for key parameters such as the droplet's spreading speed and bubble encapsulation times in both spherical and deformed regimes. By combining simulations, experimental observations, and theoretical modeling, this work aims to provide a comprehensive understanding of droplet spreading and encapsulation dynamics over the bubble's interface, paving the way for future applications and advancements in the field.

## 2. Numerical and experimental methods

### 2.1. *Numerical model: ternary MRT-LBM*

Phase-field-based Lattice Boltzmann models (PF-LBM) have proven to be highly effective in simulating bubble dynamics, droplet collision, coalescence, and the complex interfacial and topological deformations involved in wetting dynamics (Amaya-Bower & Lee 2010; Chen *et al.* 2019a; Komrakova *et al.* 2013; Li *et al.* 2016; Petersen & Brinkerhoff 2021; Premnath & Abraham 2005; Shu & Yang 2013; Wang *et al.* 2019). The MRT-LBM adopted here to investigate the bubble-droplet interaction is the extension of our previously proposed three-phase SRT-LBM (Ebadi & Hosseinalipour 2022; Kalantarpour *et al.* 2020). The implementation of multi-relaxation time into the model improved the stability of the simulation at higher density ratios [O(1000)] between gas and liquid phases, leading to the enhancement of the model's capability to handle higher velocities and extensive range of viscosities. Simulation of the three-component system is carried out in one-fluid context and mixture theory, where continuity and Navier-Stokes (NS) equations govern fluid flow and momentum exchange, and the Cahn-Hilliard (CH) equation is used for interface capturing between three immiscible components ($\phi_1$, $\phi_2$, $\phi_3$) (Boyer & Lapuerta 2006):

$$\boldsymbol{\nabla}.\mathbf{u} = 0 \tag{2.1}$$

$$\rho\left(\frac{\partial \mathbf{u}}{\partial t} + \mathbf{u}.\boldsymbol{\nabla}\mathbf{u}\right) = -\boldsymbol{\nabla}p + \boldsymbol{\nabla}.\left[\nu\rho(\boldsymbol{\nabla}\mathbf{u} + \boldsymbol{\nabla}\mathbf{u}^T)\right] + \mathbf{F}_s + \mathbf{F}_b \tag{2.2}$$

$$\frac{\partial \phi_i}{\partial t} + \boldsymbol{\nabla}\cdot\phi_i\mathbf{u} = \boldsymbol{\nabla}\cdot\left(\frac{M_i}{\lambda_i}\boldsymbol{\nabla}\mu_i\right) \tag{2.3}$$

$$\mu_i = \frac{4\lambda_r}{\zeta}\sum_{j\neq i}\left[\frac{1}{\lambda_j}\left(\frac{\partial F}{\partial \phi_i} - \frac{\partial F}{\partial \phi_j}\right)\right] - \frac{3}{4}\zeta\lambda_i\nabla^2\phi_i, i = 1, 2, 3 \tag{2.4}$$

In the above equations, **u** is the mean velocity of the ternary mixture, $p$ is hydrodynamic pressure, $\rho$, and $v$ are the density and kinematic viscosity of the mixture respectively. $\mathbf{F}_b$ Accounts for body forces, which is mainly gravity here, and $\mathbf{F}_s$ is the surface tension force that couples NS to the CH equations. In this paper the surface tension force is calculated using the following potential form:

$$\mathbf{F}_s = \sum_{i=1}^{3} \mu_i \nabla \phi_i \tag{2.5}$$

In (2.3), $M_i$ is the mobility parameter, and $\mu_i$ is the chemical potential. For an incompressible-immiscible three-component system three index functions $\phi_1, \phi_2,$ and $\phi_3$ known as order parameters are related through the following conservation constraint

$$\sum_{i=1}^{3} \phi_i = 1, 0 \leq \phi_i \leq 1 \tag{2.6}$$

To determine the spreading conditions of the system some additional coefficients, namely $\lambda_i$ are needed. These physical parameters are defined based on the interfacial tensions between the three components:

$$\lambda_i = \sigma_{ij} + \sigma_{ik} - \sigma_{jk} \quad (i, j, k = 1, 2, 3 \text{ and } j \neq i \neq k)$$
$$\begin{cases} \lambda_1\lambda_2 + \lambda_1\lambda_3 + \lambda_2\lambda_3 > 0, \\ \lambda_i + \lambda_j > 0 \text{ for } i \neq j. \end{cases} \tag{2.7}$$

The coefficient $\lambda_T$ is calculated from $3/\lambda_T = \sum_{i=1}^{3} 1/\lambda_i$. The evolution of this ternary flow is driven by the gradient of the total free energy of the system(Boyer *et al.* 2010):

$$TFE = \int_{\Omega} \left[ \frac{12}{\zeta} F(\phi_1, \phi_2, \phi_3) + \sum_{i=1}^{3} \frac{3}{8} \zeta \lambda_i \left| \nabla \phi_i \right|^2 \right] d\Omega \tag{2.8}$$

$$F(\phi_1, \phi_2, \phi_3) = \sigma_{12}\phi_1^2\phi_2^2 + \sigma_{13}\phi_1^2\phi_3^2 + \sigma_{23}\phi_2^2\phi_3^2 \tag{2.9}$$
$$+ \phi_1\phi_2\phi_3 \left[ \lambda_1\phi_1 + \lambda_2\phi_2 + \lambda_3\phi_3 \right] + \Lambda \phi_1^2\phi_2^2\phi_3^2$$

Here, $\Omega$ is the area that encloses the flow field domain, $F(\phi_1, \phi_2, \phi_3)$ is the bulk free energy, which has a triple-well structure for a three-phase system. The gradient terms on the right hand of (2.8) are capillary terms that account for interfacial energy between the phases. $\zeta$ is a characteristic parameter relating to the interface thickness, $\sigma_{ij}$ is the interfacial tensions between components $i$ and $j$, and $\Lambda \geq 0$ is a constant. The last term in (2.9) is a mathematical term added to the energy potential to prevent the bulk free energy from becoming negative, thus enabling the model to simulate total wetting states, provided that $\Lambda$ is chosen large enough.

To extend our three-phase SRT model (Kalantarpour *et al.* 2020) to MRT-LBM, we use the modeling framework of Liang *et al.*(2014), which has been developed for two-phase flows and proven by the authors that can recover correct NS and CH system of equations (2.1)-(2.3) through Chapman-Enskog (CE) analysis. In the three-phase setting LB equation with the MRT collision operator for CH and NS can be written as:

$$f_k^i(\mathbf{x}+\mathbf{e_k}\delta_t, t+\delta_t) - f_k^i(\mathbf{x},t) = -[\mathbf{M}^{-1}\mathbf{S}_f^i\mathbf{M}]_{kl}\left[ f_l^i(\mathbf{x},t) - f_l^{i,eq}(\mathbf{x},t) \right] + \delta_t \left[ \mathbf{M}^{-1}(\mathbf{I}-\frac{\mathbf{S}_f^i}{2})\mathbf{M} \right]_{kl} F_k(\mathbf{x},t) \qquad (2.10)$$

$$g_k(\mathbf{x}+\mathbf{e_k}\delta_t, t+\delta_t) - g_k(\mathbf{x},t) = -[\mathbf{M}^{-1}\mathbf{S}_g\mathbf{M}]_{kl}\left[ g_k(\mathbf{x},t) - g_k^{eq}(\mathbf{x},t) \right] + \delta_t \left[ \mathbf{M}^{-1}(\mathbf{I}-\frac{\mathbf{S}_g}{2})\mathbf{M} \right]_{kl} G_k(\mathbf{x},t) \qquad (2.11)$$

Here, $f_k^i(\mathbf{x},t)$ and $f_k^{i,eq}(\mathbf{x},t)$ represent the particle and equilibrium distribution function for the order parameters $\phi_i$, respectively. $\mathbf{M}$ is an orthogonal matrix used to transform distribution functions from lattice to moment space, and $\mathbf{M}^{-1}$ is its inverse matrix employed to transfer back the projected quantities(Krüger *et al.* 2017). $\mathbf{I}$ is the identity matrix, $\mathbf{S}_f^i$ and $\mathbf{S}_g$ are diagonal matrixes containing relaxation times (Liang *et al.* 2014). These matrixes are given in supplementary material 1. The phase index of the third component and its chemical potential can be calculated from constraint (2.6) as $\phi_3 = 1 - (\phi_1 + \phi_2)$ and $\sum_{i=1}^3 \frac{\mu_i}{\lambda_i} = 0$, hence only components 1 and 2 need to be numerically solved.

Thus, the superscript $i$ (2.10) is either 1 or 2. $\tau_i$ and $\tau_g$ are relaxation times towards equilibrium, $\delta_t$ is the time step, $F_k^i(\mathbf{x},t)$ is the source term, and $G_k(\mathbf{x},t)$ is the total force distribution function given as (Liang *et al.* 2016):

$$F_k^i = \left(1-\frac{1}{2\tau_i}\right)\frac{\omega_k \mathbf{e_k}.\partial_t \phi_i \mathbf{u}}{c_s^2} \qquad (2.12)$$

$$G_k = \left(1-\frac{1}{2\tau_g}\right)(\mathbf{e_k}-\mathbf{u}).\left[ s_k(\mathbf{u})\nabla\rho + (s_k(\mathbf{u})+\omega_k)\frac{(\mathbf{F}_s+\mathbf{F}_b)}{c_s^2} \right] + \frac{\omega_k \mathbf{e_k}.\mathbf{F}_a}{c_s^2} \qquad (2.13)$$

$$\mathbf{F}_a = \mathbf{u}[(\rho_1-\rho_3)\nabla\cdot(M_1\nabla\mu_1) + (\rho_2-\rho_3)\nabla\cdot(M_2\nabla\mu_2)] \qquad (2.14)$$

Here $\mathbf{F}_s$ is the surface tension force given by (2.5), $\mathbf{F}_b = -\rho g\nabla H$ is the gravitational force, and $\mathbf{F}_a$ is an interfacial force introduced to recover correct momentum equation, $c_s = c/\sqrt{3}$ is the lattice speed of sound, where $c = \frac{\delta_x}{\delta_t} = 1$.

Equilibrium distribution functions $f_k^{i,eq}$ and $g_k^{eq}$ are derived from Maxwell-Boltzmann distribution using either Taylor or Hermit series expansion as (Krüger *et al.* 2017):

$$f_k^{i,eq} = \begin{cases} \phi_i + (\omega_k - 1)\eta\mu_i, \, k = 0 \\ \omega_k \eta\mu_i + \omega_k \phi_i \dfrac{\mathbf{e_k}.\mathbf{u}}{c_s^2}, \, k \neq 0 \end{cases} \tag{2.15}$$

$$g_k^{eq} = \begin{cases} \dfrac{p}{c_s^2}(\omega_k - 1) + \rho s_k(\mathbf{u}), \, k = 0, \\ \dfrac{p}{c_s^2}\omega_k + \rho s_k(\mathbf{u}), \quad k \neq 0. \end{cases}, \qquad \text{where} \quad s_k(\mathbf{u}) = \omega_k \left[ \dfrac{\mathbf{e_k}.\mathbf{u}}{c_s^2} + \dfrac{(\mathbf{e_k}.\mathbf{u})^2}{2c_s^4} - \dfrac{\mathbf{u}.\mathbf{u}}{2c_s^2} \right] \tag{2.16}$$

In these relations, $\mathbf{u}$ accounts for the macroscopic fluid velocity, $\eta$ is a tunable parameter controlling mobility $M_i$, $\mathbf{e_k}$ and $\omega_k$ are discrete mesoscopic velocities and weighting factors in $k$th direction defined as below for a D2Q9 lattice model:

$$\omega_k = \begin{cases} 4/9, \quad k = 0; \\ 1/9, \quad k = 1-4; \\ 1/36, \quad k = 5-8. \end{cases} \quad \text{and,} \quad \mathbf{e}_k = \begin{cases} (0,0)c, \quad k = 0; \\ (\cos\theta_k, \sin\theta_k)c, \quad \theta_k = (k-1)\pi/2, \quad k = 1-4; \\ \sqrt{2}(\cos\theta_k, \sin\theta_k)c, \quad \theta_k = (k-5)\pi/2 + \pi/4, \quad k = 5-8. \end{cases} \tag{2.17}$$

Macroscopic properties are obtained using moments of the distribution functions. Zeroth moment of $f_k^i$ provides phase-field variables for components 1 and 2:

$$\phi_i = \sum_k f_k^i \tag{2.18}$$

And from the conservation constraint (2.6): $\phi_3 = 1 - \phi_1 - \phi_2$. Similarly, macroscopic pressure $p$ and mixture velocity u are calculated as (Kalantarpour *et al.* 2020):

$$\mathbf{u} = \frac{1}{\rho} \left[ \sum_k \mathbf{e}_k g_k + 0.5\delta_t (\mathbf{F}_s + \mathbf{F}_b) \right] \tag{2.19}$$

$$p = \frac{c_s^2}{1 - \omega_0} \left[ \sum_{k \neq 0} g_k + \frac{\delta_t}{2} \mathbf{u}.\nabla\rho + \rho s_0(\mathbf{u}) \right] \tag{2.20}$$

Mixture density $\rho$, kinematic viscosity $\nu$, and mobility parameter are calculated using linear interpolation and CE analysis:

$$\rho = \sum_{i=1}^{3} \rho_i \phi_i \tag{2.21}$$

$$M_i = c_s^2 (\tau_i - 0.5)\delta_t \tag{2.22}$$

$$\nu = c_s^2 (\tau_g - 0.5)\delta_t \tag{2.23}$$

where,

$$\tau_g = 0.5 + \sum_{i=1}^{3} \phi_i \left( \tau_i - 0.5 \right) \tag{2.24}$$

$$\tau_i = \frac{\nu_i}{c_s^2 \delta t} + 0.5, \qquad i = 1, 2, 3. \tag{2.25}$$

Here $\nu_i$ is the kinematic viscosity of components 1, 2, and 3, which are set to a constant value prior to the simulations.

## 2.2. Simulation setup

The schematic configuration of the interaction phenomenon between rising bubble-droplet in a continuous liquid medium is depicted in Figure 1. The air bubble (denoted as phase 3) is placed at a fixed distance of $Y_{cb}$=0.3$H$-1.5$R$ from the bottom of the rectangular domain, and the oil droplet (represented by phase 2) is initially located at $Y_{cd}$=0.3$H$+1.5$R$, where $R$=$R_d$=$R_b$=60 lattice unit (l.u.) for the equal sized bubble-droplet, which is equivalent of 0.4 mm in real world. For the investigation of the size effect, bubble and droplet locations are set using $R_d$ and $R_b$ in a way that the separation distance always equals 3$R$. Gravitational acceleration ($g$) is applied to the whole domain, and both bubble and droplet are initially at rest and start to ascend under the action of the buoyancy force. The initial distributions of the order parameters are defined as:

$$\phi_2(x, y) = 0.5 + 0.5 \tanh \left( \frac{2}{\zeta} (R - \sqrt{(x - x_{c1})^2 + (y - y_{c1})^2}) \right)$$
$$\phi_3(x, y) = 0.5 + 0.5 \tanh \left( \frac{2}{\zeta} (R - \sqrt{(x - x_{c2})^2 + (y - y_{c2})^2}) \right) \tag{2.26}$$
$$\phi_1(x, y) = 1 - (\phi_2(x, y) + \phi_3(x, y))$$

In order to convert the LBM parameters to physical space we use the suggested procedure in (Komrakova *et al.* 2013; Krüger *et al.* 2017), where three independent conversion factors defined as $C_\psi = \frac{\psi_{physical}}{\psi_{lattice}}$ are employed to map the two space units and to obtain the remaining relevant quantities. Here, $\psi$ can represent any quantity or parameter. We adopt density, surface tension, and viscosity as our independent conversion factors and set them as $C_\rho = 1000$ kg/m³, $C_\sigma = 2.94$ N/m, $C_\nu = 0.00014$ m²/s. Using dimensional analysis, the conversion factors for the length, time, velocity, and gravity scales are respectively calculated as $C_l = 6.6667 \times 10^{-6}$ m, $C_t = 3.175 \times 10^{-7}$ s, $C_u = 21$ m/s, $C_g = 66150000$ m/s². For the full description of the unit conversion procedure see supplementary material 3. Based on our pre-analysis and evaluation of the models in terms of the acceptable range of parameters we have chosen $\hat{\rho}_1 = 1$, $\hat{\rho}_2 = 0.85$, $\hat{\rho}_3 = 0.001$, $\hat{v}_1 = 0.01$, $\hat{v}_2 = 0.0714$, $\hat{v}_3 = 0.108$, and $\hat{\sigma}_{12} = 0.01$, $\hat{\sigma}_{13} = 0.0196$, $\hat{\sigma}_{23} = 0.009$ as the base values for our simulations. With this set of lattice values and conversion factors, the continuous phase can be regarded as water, the bubble phase as air, and the droplet phase as a typical oil relevant to practical applications such as gas flotation. Thus, their physical properties are considered as $\rho_1 = 1000$, $\rho_2 = 850$, $\rho_3 = 1.$ (kg/m$^3$), $v_1 = 1.4 \times 10^{-6}$, $v_2 = 10 \times 10^{-6}$, $v_3 = 1.52 \times 10^{-5}$ (m$^2$/s), and $\sigma_{12} = 0.0294$, $\sigma_{13} = 0.0576$, $\sigma_{23} = 0.0265$ (N/m). It should be noted that to allow the proposed MRT-LB model to simulate wider ranges of interfacial tensions (IFTs) and viscosity ratios, some compromises have been made in the real values of water viscosity and its IFT with air. The dimensionless numbers are defined as:

$$S_2 = \sigma_{13} - (\sigma_{12} + \sigma_{23}) \tag{2.27}$$

$$Oh_s = \frac{\mu_{1,2}}{\sqrt{\rho_{1,2} R_d S_2}} \tag{2.28}$$

$$Ca_s = \frac{\mu_{1,2} \sqrt{g R_d}}{S_2} \tag{2.29}$$

$$\mathrm{Re}_{d,b} = \frac{\sqrt{g} \times (2R_{d,b})^{3/2}}{v_1} \tag{2.30}$$

$$Bo_{d,b} = \frac{4g(\rho_1 - \rho_{2,3})(R_{d,b})^2}{\sigma_{12,13}} \tag{2.31}$$

$$Mo_{d,b} = \frac{g(\rho_1 - \rho_{2,3})(v_1)^4(\rho_1)^2}{(\sigma_{12,13})^3} \tag{2.32}$$

$$B = \frac{b}{2R} \tag{2.33}$$

In the above relations $Oh$, $Ca$, $Re$, $Mo$, and $Bo$ are the Ohnesorge, Capillary, Reynolds, Morton, and Bond otherwise known as Eötvös number, respectively. $B$ is the collision parameter, used to set the horizontal distance of the bubble-droplets to each other in off-center interactions. Subscripts $d$ and $b$ represent drop (phase 2) and bubble (phase 3), respectively. Subscripts 1 in $Oh_s$ and $Ca_s$ represent the cases where continuous phase properties have been used (mainly in the investigation of the droplet and substrate fluid viscosity effect). The spreading coefficient $S_2$ is used to account for the three-component nature of the interaction (represented by index s). For the size effect investigation, the value of the parameter that is being varied, $R_d$ or $R_b$ is chosen as the length scale. Also, since bubble/droplet rising speed is not a priory known, the gravitational characteristic velocity is calculated as $\sqrt{g(2R_{d,b})}$. Therefore, in the analysis of the results adopting $Oh$ and $Bo$ numbers will be preferred over $Re$ and $Ca$. In all the numerical tests of this work, periodic boundary conditions have been applied to the left and right sides of the domain, while bounce-back scheme has been used for the top and bottom walls. The interface thickness is $\xi$=4 lattice units, and other parameters are considered as: $\Lambda$=0.007, $M$=0.005. Based on our analysis of the domain size influence (see supplementary materials 2) a mesh of 7R*14R is used for equal-sized interactions, while 8R*16R is adopted for the investigation of size effect. These grid dimensions are consistent with previous studies on bubble-droplet dynamics (Amaya-Bower & Lee 2010; Komrakova *et al.* 2013; Zhao & Lee 2023) and considering a relatively high number of simulations they produce a reasonable accuracy. Also, the bubble and droplet are resolved with at least $D$=120 lattice units ($R$=60 l.u), which is well above the customary size of $R$ or even $D$=30 l.u. adopted in most LBM studies (Amaya-Bower & Lee 2010; Komrakova *et al.* 2013; Zhao & Lee 2023).

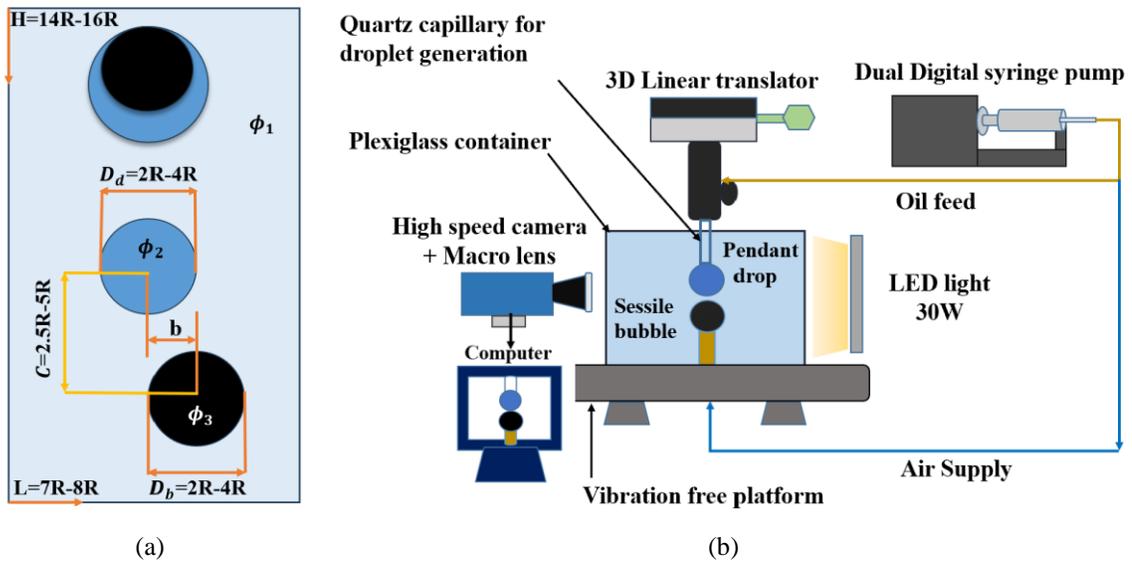

(a)                                                                                            (b)

Figure 1. Configuration and parameters of drop-bubble rise and collision in the presence of gravity field (a), and schematic of the experimental setup for quasi-static interaction between a pendant droplet and sessile air bubble in deionized water (b). the substrate fluid, i.e. water (phase 1-$\phi_1$), the droplet (phase 2- $\phi_2$), and the bubble (phase 3- $\phi_3$) are represented by light blue, dark blue, and black colors respectively. To minimize surface tension effects on droplet formation quartz capillaries with an inner diameter (I.D.) of 0.375 mm were used for drop generation, and copper needles with I.D. ~ 0.1 mm were employed for bubble injection. Droplets and bubbles were produced via an automatic syringe pump in the range of $D_{d,b}$=1-2.5 mm diameters. In the simulation domain, $R$=60 lattice units (l.u), equivalent to 0.4 mm in physical space.

## 2.3. *Experimental set-up*

A schematic representation of the experimental apparatus used to observe the bubble-droplet interaction is shown in Figure 1(b). The setup, consists of a rectangular container with dimensions of 14 * 15 * 6 cm³ filled with de-ionized water (DIW). Vertical alignment of the drop (upper) and bubble (lower) generation nozzles allows us to probe the bubble-droplet interactions in both quasi-static and dynamic states. In quasi-static interaction, bubbles and droplets are grown in size using an automatic syringe pump until apparent contact is formed. Then the pump is stopped to let the film drainage start by the action of capillary and surface forces. In the dynamic case, the sessile bubble detaches from the needle, rises under buoyancy force, and collides with the pendant droplet. Since we are interested in the droplet spreading process rather than film drainage, and also to keep the experiments consistent with simulation settings, the separation distance between the two nozzles was set to 10 mm in the dynamic case. PCO-1200 HS high-speed camera integrated with a Carl Zeiss 1X macro lens has been used to record the process. The images are recorded at 2000-4000 frames per second (fps) rate. The materials used in the experiments and their physical properties are given in Table 1.

| Fluids | $\rho$ (kg/m$^3$) | $\mu$ (mPa.s) | $\sigma_{wa}$ (mN/m) | $\sigma_{wo}$ (mN/m) | $\sigma_{oa}$ (mN/m) | $S_o$ (mN/m) |
|---|---|---|---|---|---|---|
| Air | 1.2 | 0.0185 | N / A | N / A | N / A | N / A |
| DI water | 998 | 0.89 | 72.44 | N / A | N / A | N / A |
| Gasoline | 728 | 0.62 | N / A | 29.5 | 22.508 | 20.43 |
| Sunflower oil | 906 | 58 | N / A | 21.93 | 27.217 | 23.30 |
| Silicone oil 10 | 941 | 10.3 | N / A | 35.13 | 18.842 | 18.48 |

Table 1. Specifications of the air, water, and oil droplets used in the experiments.

## 2.4. *Model validation*

The base SRT-LB model used to derive the proposed ternary MRT-LBM has been extensively validated through benchmark tests in our previous publications, including two and three-component Rayleigh-Taylor instability, ternary lens spreading, passage of bubble through immiscible interfaces, in Kalantarpour et al. (2020), domain size influence and grid dependency in droplet collision, coalescence of miscible and immiscible droplets, double emulsion configuration and possible droplet engulfment morphologies inEbadi & Hosseinalipour (2022). Hence, here we only perform two application-based test cases of terminal velocity of rising droplets, and droplet encapsulating a bubble to verify the MRT extension with experimental, theoretical, and numerical data to showcase the capability of the model in handling encapsulation phenomena and its consistency with physical values.

| Liquid system | $\rho$ (kg/m$^3$) | $\mu$ (mPa.s) | $\sigma_{wo}$ (mN/m) | Mo |
|---|---|---|---|---|
| **Toluene** **Water** | 862.3 997.2 | 0.552 0.89 | 35 | $1.95 \times 10^{-11}$ |
| ***n*-butanol** **Water** | 845.4 986.5 | 3.28 1.39 | 1.63 | $1.23 \times 10^{-6}$ |

Table 2. Physical properties of the materials used for the drop rise velocity problem.

### 2.4.1. *Droplet terminal velocity*

Configuration of the problem is the same as Figure 1 (a), where $\phi_3$=0, and $\phi_1$ and $\phi_2$ are initialized according to (2.26). Since we are interested in terminal rising velocity, the domain size in this test case is set to 7$R$*24$R$, where $R$=30-60 l.u. is the drop radius. The experimental data for the physical properties of toluene/water (Wegener *et al.* 2010) and $n$-butanol/water (Bertakis *et al.* 2010) systems are given in Table 2. In order to match these properties in LBM simulations $C_\rho = 997.2$ kg/m$^3$, $C_\sigma = 0.29 - 0.5$ N/m, $C_\nu = 0.00009$ m$^2$/s, $C_l = 1.6 - 2.8 \times 10^{-5}$ m, $C_u = 3.22 - 5.57$ m/s for the former (toluene), and $C_\rho = 32.88$ kg/m$^3$, $C_\sigma = 0.033 - 0.04$ N/m, $C_\nu = 0.000138 - 0.0002$ m$^2$/s, $C_l = 1.5 - 3 \times 10^{-5}$ m, $C_u = 5.75 - 8.99$ m/s for the latter has been used. The simulation results for the terminal rise velocity at different drop diameters in the range of 0.5-3 mm for toluene and between 1-2.48 mm for $n$-butanol droplets are given in Figure 2 (a) and (b) respectively. As can be seen for both droplet types the agreement between the present LBM simulations and experimental data of (Bertakis *et al.* 2010; Wegener *et al.* 2010) and numerical results of (Bäumler *et al.* 2011; Komrakova *et al.* 2013) is very good. The results also fit the empirical correlation proposed by Henscke Bäumler et al. (2011) for both spherical and deformed regimes that are relevant to the scope of this paper.

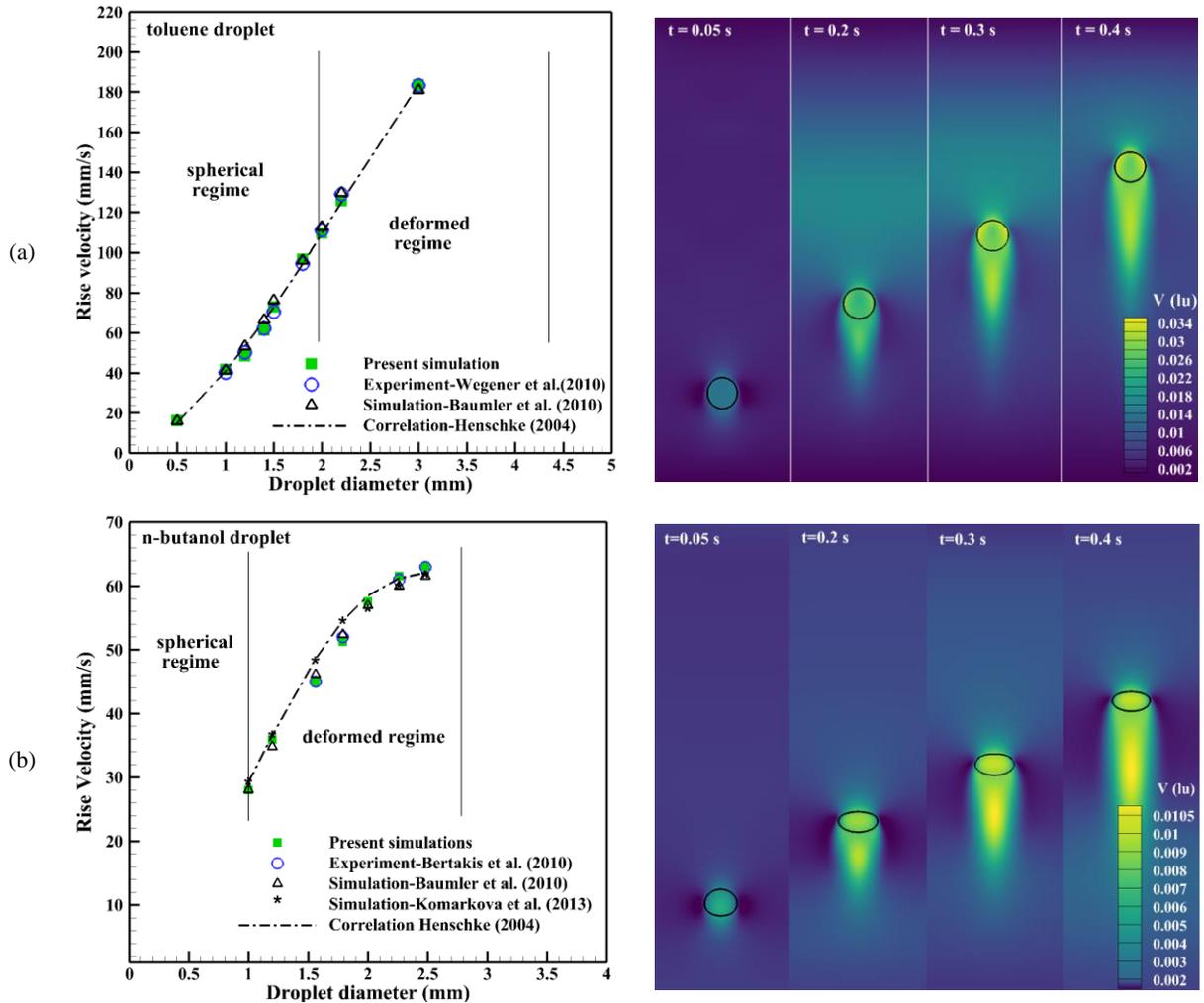

Figure 2. Terminal velocity of rising droplet versus drop diameter (on the right) and temporal evolution of the velocity contours for 2 mm drop rising in water (on the right), panel (a) and (b) represent toluene and $n$-butanol droplets, respectively. Legends in velocity contours are in lattice units, to convert these values to physical units they should be multiplied by $C_u = 3.22$ m/s for toluene and $C_u = 5.75$ m/s for *n*-butanol which would give the terminal velocity of 109.48 mm/s and 60 mm/s, respectively.

Velocity contour for a 2 mm toluene and *n*-butanol droplet at various time strands from 0.05 s to 0.4 s that the drop reaches its maximum rise velocity, i.e. terminal velocity is displayed on the right side of Figure 2. (a) and (b), respectively. It can be seen that for toluene drop, 2 mm is the borderline for change from spherical to deformed regime, as also visible from slight drop shape deformation in t=0.4 s, while the 2 mm *n*-butanol droplet owing to its lower surface tension is completely in the deformed regime, leading to higher drag force on the drop and lower rising velocity. The droplet shape deformations and the wake behind the butanol drop are consistent with the numerical results of Komrakova et al. (2013).

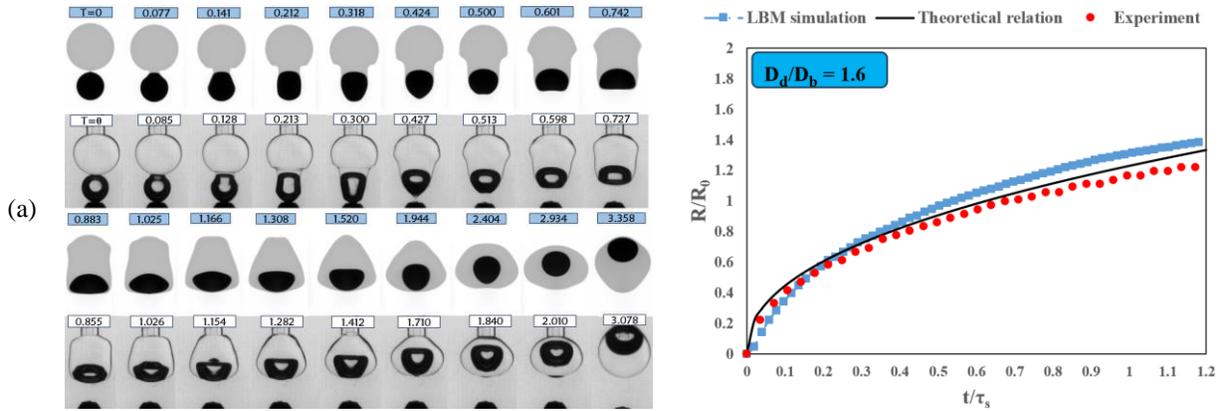

(For the caption see Figure 3.)

### 2.4.2. *Quasi-static bubble-droplet interaction*

In this test, the LBM simulations are validated by quasi-static experiments where the initial velocity of the interaction is negligible. In the experiments, the bubble and droplet are grown to form an apparent contact, then the pump is stopped to allow for spontaneous film drainage and rupture. The simulations are initiated at the contact state, i.e. center to center distance of the bubble and droplet is set to $2.1R$, thus the film drainage stage is very short and negligible and the interaction starts with the droplet spreading. Unfortunately, due to the numerical limitations, we couldn't produce exact bubble-droplet sizes in the simulations. Instead, to reproduce similar behavior in both spaces we have used dimensionless parameters $Bo_b$=0.3, 0.71, 0.4 , $Bo_d$=0.51, 0.47, 0.16 and $Oh_s$=0.0046, 0.0048, 0.0063 with $D_d/D_b$=1.6, 1, 0.78, respectively to match the experimental conditions. The lattice conversion factors are calculated in the same manner as described in the simulation setup section.

The sequences of the encapsulation process with dimensionless time T = $t / \tau_{ic}$ are displayed in Figure 3 and compared with experimental images. Here, $\tau_{ic}$ is the capillary-inertial time scale defined using the spreading coefficient instead of a single surface tension (see caption of the figure). It can be seen that the LBM results are in general agreement with experimental observations for the shape evolution of the bubble-drop compound during liquid bridge growth, encapsulation process, as well as propagation and convergence of the capillary waves. The neck growth rate with T is depicted in front of each $D_d/D_b$ case on the right side of Figure 3. For the droplets larger than, or equal to bubble size the growth rate follows the power-law approximation proposed by Li et al. (2014) for the passage of a small bubble through a curved liquid-liquid interface:

$$R / R_0 = 1.23(t / \tau_{IC})^{0.44} \tag{2.34}$$

For the droplet smaller than the bubble ($D_d/D_b$=0.78) the liquid bridge growth obeys a power-law with the exponent of 1/2 reported for coalescence of bubbles, miscible (Anthony *et al.* 2017; Paulsen *et al.* 2014) or immiscible droplets (Xu *et al.* 2024). In this case, the bridge evolves as

$$R / R_0 = 1.3(t / \tau_{IC})^{0.5} \tag{2.35}$$

The maximum difference in estimating neck growth $R/R_0$ is %12.5 for $D_d/D_b$=1, which seems reasonable since the bubble and droplet are formed at the tip of the needle in the experiments.

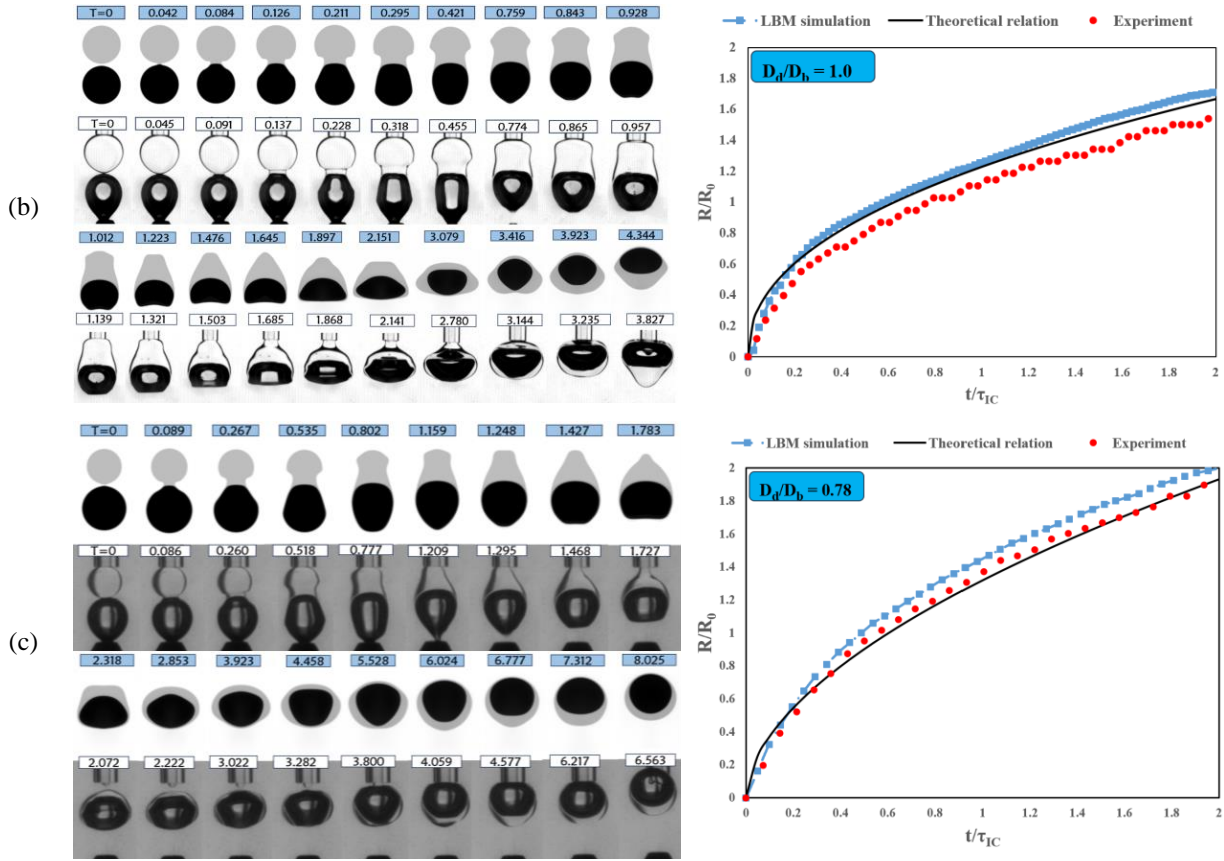

Figure 3. Comparison of the experimental sequences of a gasoline droplet encapsulating an air bubble in a quasi-static state (lower rows on the left) and numerical simulations of the problem using MRT-LBM (upper rows on the left). Profiles on the right depict the normalized neck growth rate or more specifically horizontal stretching of the droplet during the spreading for droplet-to-bubble size ratios of (a) $R_d$=1.2 mm, $R_b$=0.75 mm, (b) $R_d$=$R_b$=1.15 mm and, (c) $R_d$=0.66 mm, $R_b$=0.85 mm. The neck growth rate $R$ is normalized by the initial drop radius for each case. Here, the inertial-capillary time scale is defined as $\tau_{ic} = \sqrt{\rho_1 R_s^3 / S_2}$ using the spreading coefficient of the oil drop, thus in the above time strands $T = t / \tau_{ic}$. As a scale bar, the size of the drop generation needle is 1mm.

Like the coalescence phenomena, the spreading of the droplet over the bubble's interface is driven by the release of surface energy which is associated with the capillary pressure of the drop. For low-viscosity fluids, Thoroddsen *et. al.* (Thoroddsen *et al.* 2005b) derived a simple model for the outward velocity of the neck by balancing capillary and dynamic pressures ($\rho u^2$):

$$\frac{\mathrm{d}R}{\mathrm{d}t} = C \sqrt{\frac{\sigma}{\rho}\left(\frac{1}{\delta} - \frac{1}{R}\right)} \qquad (2.36)$$

In which $C$ is a constant, $R$ is the radial location of the neck, and $\delta$ is the vertical distance between the drop's surfaces at the instant location of $R$. For two unequal-sized drops with the radius of $R_1$ and $R_2$, Deka et al. (2019) calculated $\delta$ as $\delta = R_1 + R_2 - \sqrt{R_1^2 - R^2} - \sqrt{R_2^2 - R^2}$. Based on these relationships it can be deduced that the expansion rate of the neck is higher for smaller drops, leading to a higher slope in the neck growth curve for $D_d/D_b$=0.78 (Figure 3 (c)) compared to the two other cases in (a) and (b) with bigger drop sizes. It is also observed that the droplet stretches to a higher degree for case (c), reaching two times its initial radius ($R/R_0$=2), in contrast to 1.3 and 1.6 in the former two, respectively. The effect of bubble-droplet size on the spreading speed of the drop and encapsulation along with the maximal stretching of the compound during this phenomenon will be discussed in greater depth in the size effect investigation in §3.5.

## 3. Results and discussion

The results of the simulations will be presented in two categories. First, we will investigate the effect of changing physical properties of the droplet phase and substrate fluid on the interaction of the equal-sized bubble-droplet. For this category, which will be denoted as the spherical regime ($g=g_0$=9.81 m/s$^2$), based on the parameters range given in Table 3, the smallest capillary length for the bubble and droplet are $\lambda_{cb} = \sqrt{\sigma_{13}/\Delta\rho_{13}g} = 2.42\,\mathrm{mm}$, $\lambda_{cd} = \sqrt{\sigma_{12}/\Delta\rho_{12}g} = 3.47\,\mathrm{mm}$, respectively. Hence, we can ignore gravitational effects on the encapsulation process, since the bubble and droplet shapes remain spherical prior to the contact. It should be noted that in the bubble and droplet velocity profiles the average vertical ($v$) and horizontal ($u$) velocities are obtained using the following averaging relations $\bar{X}_i = \sum \rho \phi_i x / \sum \rho \phi_i$, where $x$ can be $u$ or $v$, and $i$ represents the component that is being averaged (2 and 3 for the drop and bubble phases respectively), $\phi$ is the order parameter and $\rho$ is the density.

| variable | $\rho_2$ (kg/m$^3$) | $\nu_2$ (cSt) | $\nu_1$ (cSt) | $\sigma_{12}$ (mN/m) | $\sigma_{23}$ (mN/m) | $S_o$ (mN/m) | $Bo_b$ | $Bo_d$ | $Oh_S$ |
|---|---|---|---|---|---|---|---|---|---|
| Base case (SS0) | 850 | 10 | 1.4 | 29.4 | 26.5 | 1.47 | 0.11 | 0.032 | 0.38 |
| Drop vis. (SS1) | 850 | 0.98-56 | 1.4 | 29.4 | 26.5 | 1.47 | 0.11 | 0.032 | 0.037-2.13 |
| Medium vis. (SS2) | 850 | 10 | 1.4-56 | 29.4 | 26.5 | 1.47 | 0.36 | 0.105 | 0.28-2.3 |
| IFT (SS3) | 850 | 10 | 1.4 | 17.6-50 | 17.6-29.5 | 1.47-10.3 | 0.11 | 0.023-0.053 | 0.14-0.85 |
| Density (SS4) | 700-950 | 10 | 1.4 | 29.4 | 26.5 | 1.47 | 0.11 | 0.01-0.064 | 0.38-0.42 |

Table 3. The range of physical parameters and important non-dimensional numbers used for the simulation of equal-sized bubble-droplet interaction. Here, SS is short for simulation series. Physical properties of the base case were given in §2.2. Based on these properties, $Mo_b = 2\times10^{-10}$, $Re_d = Re_b = 50.57$, $Mo_d = 7.4\times10^{-11} - 1.03\times10^{-9}$ and $Ca_s = 0.05 - 2.86$.

### 3.1. The base case (SS0)

In the simulation of the base case, used as a reference for changing the physical parameters, both the bubble and droplet have equal radii of $R$=400 µm (60 l.u.). The physical characteristics for this case are provided under SS0 of Table 3. Figure 4 presents typical sequences of bubble–droplet interaction in a dynamic (collisional) state and includes a qualitative comparison with experimental images of a silicone oil droplet interacting with an air bubble of approximately the same size ($R_d \approx R_b$=1.1 mm). This comparison aims to offer visual insight into the different stages of the interaction and to support the interpretation of simulation results. The time is reset to zero at the moment of physical contact such that the film drainage phase begins at $t^*$=$t$-$t_{col}$=0. As the bubble approaches the droplet interface, the interstitial liquid between them is expelled radially outward due to a rise in film pressure, which includes both capillary forces and disjoining pressure effects. In the simulation, the film drainage occurs almost instantaneously, within less than 1 ms, owing to the absence of explicit repulsive surface force models. By contrast, in the experimental observations, the drainage phase lasts approximately 4.8 ms, highlighting the effect of additional physical interactions present in the real system. Immediately after film rupture and liquid bridge formation, a sharp increase in both horizontal and vertical velocities is observed in the contact region (notably at $t$=21.58 in Figure 5 (a) and (b)). The liquid bridge then expands as the droplet spreads around the bubble, with capillary waves traveling along the three-phase contact line. These waves converge beneath the bubble, briefly forming a triangular configuration, visible at $t^*$=7.68 ms in the experimental images ($t^*$=3.49 ms in the LBM simulations). This process is followed by full encapsulation of the bubble at $t$=10.56 ms. Subsequently, the bubble reshapes within the drop over the interval $t^*$=13.44-20.64 ms, eventually detaching from the rear end of the droplet. Then the bubble gradually migrates to the front side, and leads the compound rise with a trailing thick oil layer.

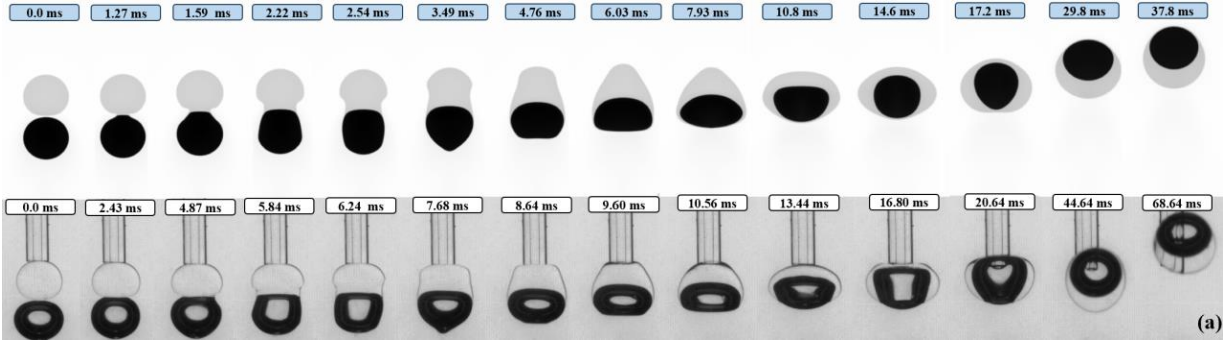

Figure 4. (a) Numerical (top) and experimental (bottom) snapshots of typical equal-sized droplet-bubble collision and the encapsulating process. Silicon oil (see Table 1) with a kinematic viscosity of $10.09\times10^{-6}$ m$^2$/s has been used in the experiments, which is very close to the viscosity of the drop in the base case. Droplet and bubble are approximately of the same size of 1.1 mm in the experiments, while $R_d=R_b=0.56$ mm in the LB simulations.

Vertical velocity profiles of the droplet and bubble shown in Figure 5 (a) provide insightful information regarding the interaction dynamics in all stages of ascend, collision, and encapsulation. Based on these curves and their comparison with the velocity contours depicted in Figure 5 (b), the whole interaction process can be divided into 4 consecutive stages: 1) The collision stage, which includes the moment the bubble and the drop start moving under the effect of buoyancy until the physical contact between the two is formed at $t_{col}=20$ ms. 2) The spreading or encapsulation stage, begins with the discharge and rupture of the intervening film and ends with the complete engulfment of the bubble at $t=26.35$ ms ($t_{enc}=6.35$ ms). 3) Bubble reshaping stage and rising inside the drop, where encapsulated bubble quickly adapts to the new viscous medium under the interfacial tension effects. This stage occurs immediately after the encapsulation of the bubble and extends to the moment when the bubble reaches the drop's top surface ($t=26.35$-47.62 ms). This stage can be divided into consecutive steps of reshaping and detachment as shown in Figure 5 (a). 4) Compound rising stage, where the bubble-drop aggregate reaches its terminal rising velocity and ascends to the surface of the pool ($t=47.62$-80 ms).

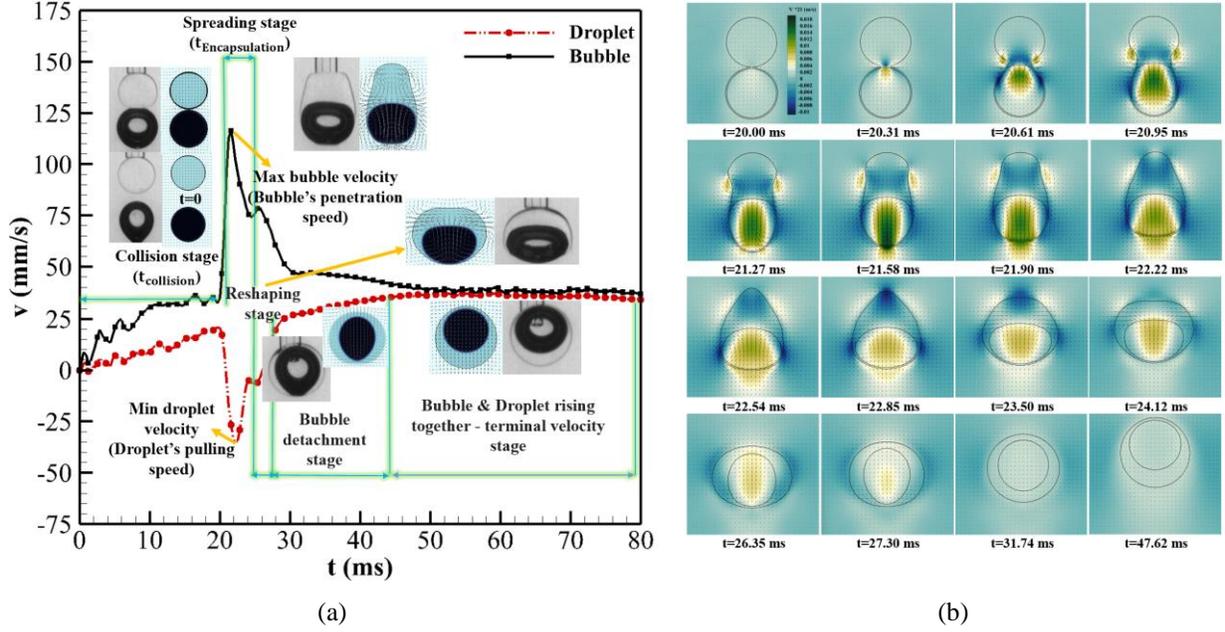

(a)

(b)

Figure 5. (a) temporal evolution of the vertical bubble/droplet velocity profiles with the identified stages during the interaction, and (b) velocity contours at various time strands from the onset of neck formation to full engulfment of the bubble for the base case ($R_d = R_b = 0.4$ mm).

As shown in Figure 5 (a), and as expected from buoyancy-driven dynamics, the bubble exhibits a higher rising velocity than the droplet during the collision stage, resulting in a relative impact velocity at the moment of contact. During the penetration phase, the bubble accelerates upward to acquire a maximum velocity of 115 mm/s, which is interestingly about 223% of its impact velocity. This point will be referred to as the maximum bubble velocity or bubble penetration speed. Concurrently, the spreading droplet is displaced downward, reaching a minimum velocity of -35 mm/s. Owing to its negative sign and the downward direction, this is termed the minimum drop velocity, or the drop's pulling speed. Both of these extremum points can be clearly identified in velocity contours of Figure 5 (a) between $t$=21.58 to 21.90 ms. The movement of capillary waves is also perfectly visible in these contours within the time frame $t$= 20.61-22.85 ms. These waves propagate along the three-phase contact lines at the bubble–droplet–pool interface (seen as blue vortical spots migrating toward the bottom of the bubble) and along the drop–pool interface (represented by light brown regions moving upward along the curved edges of the neck). Eventually, these capillary waves converge at the droplet apex, stretching it into a triangular or corner-like shape. The intensity and behavior of these capillary waves depends on the physical properties of the drop and pool, specifically viscosity and interfacial tension. For low-viscosity fluids capillary waves can lead to pinch-off or satellite formation, which has been the subject of numerous studies (Deka *et al.* 2019; Li *et al.* 2014a).

### 3.2. *Droplet and continuous phase viscosity*

The viscosities of both the droplet and the surrounding medium play a crucial role in determining droplet coalescence and spreading behavior. In this section, we examine the influence of viscous effects on the collision and encapsulation times of a rising droplet–bubble pair. We also demonstrate how encapsulation time can be quantitatively extracted from the evolution of the phase-field index. The sequential stages of bubble encapsulation by a relatively high-viscosity oil droplet are illustrated in Figure 6 (a). To enhance the visibility of encapsulation dynamics and to preserve near-spherical shapes, the experiments are conducted under quasi-static conditions. The upper row of Figure 6 (a) shows numerical results for a droplet of comparable viscosity undergoing dynamic (collisional) interaction. Compared to the lower-viscosity droplet in the baseline case, as well as gasoline droplets discussed in §2.4.2. , the high-viscosity droplet exhibits fewer surface undulations and smoother interfaces due to dampened capillary waves. This suppression is caused by viscous resistance to interfacial deformation during spreading. Notably, in the lowest-viscosity case, the droplet exhibits a sharp, corner-like tip ($t$=2.22 ms in Figure 7 (a)), whereas increasing viscosity leads to a more flattened tip profile. Given that the release of interfacial energy is consistent across all cases, the observed reduction in capillary wave propagation and amplitude is attributed to increased viscosity. This results in slower spreading (or pulling) speeds and consequently longer encapsulation times.

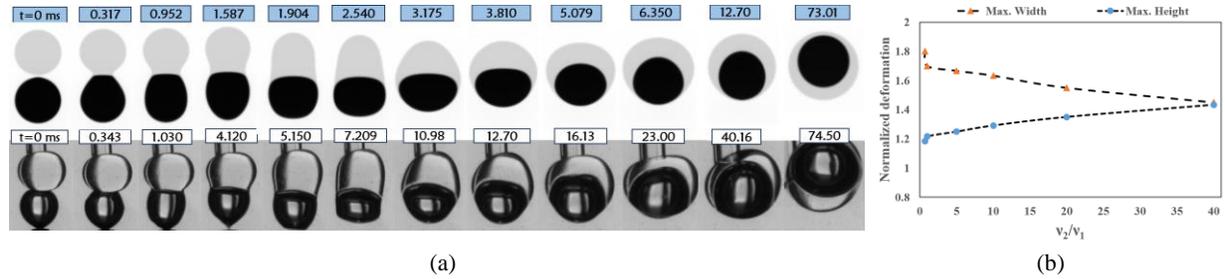

(a)                                                                (b)

Figure 6. (a) Typical numerical (top row-$Oh_d$=0.48) and experimental (bottom row-$Oh_d$=0.415) snapshots of a high-viscosity oil drop encapsulating the bubble. The drop material in the experiment is sunflower oil (see Table 1) with $R_d$=1.1 mm, and the bubble radius is $R_b$=0.95 mm. For the LBM simulation $R_d$= $R_b$=0.4 mm and physical properties are according to SS1 of Table 3 with the highest viscosity $v_{drop} = 56 \times 10^{-6}$ (m$^2$/s) . (b) Maximal horizontal and vertical deformation of the bubble-drop compound during the encapsulation process normalized by the drop's initial diameter $D_0$ at different viscosity ratios. This represents the variation of the compound's height ($d_v$) versus its width ($d_h$) with increasing drop viscosity.

Figure 6 (b) displays the maximum horizontal ($d_h$) and vertical ($d_v$) stretching of the compound during encapsulation across different drop-to-medium viscosity ratios. The greatest horizontal deformation, approximately 1.8 times the undeformed diameter, is observed at the lowest studied viscosity. In contrast, the smallest deformation (where $d_h$= $d_v$=1.4) occurs at the highest droplet viscosity. This trend indicates that, as viscosity increases, viscous stresses increasingly dominate over capillary-inertial forces, resulting in a rounder compound morphology. Consequently, the aspect ratio approaches unity ($d_h/d_v \rightarrow 1$) at higher viscosities.

The contours of the order parameter for the droplet phase $\phi_2$ for the lowest (a), the highest (c), and the base case (b), which is considered an intermediate viscosity, are depicted in Figure 7. In these contours, the droplet phase is shown in dark purple, while the bubble and surrounding fluid phases appear in light blue. At the highest viscosity, the liquid bridge grows more uniformly in both horizontal and vertical directions, accompanied by noticeably smoother interfaces and fewer surface corrugations. This behavior is consistent with experimental observations: comparing the gasoline droplets (low viscosity) in Figure 3 with the sunflower oil droplets (high viscosity) in Figure 6 (a) confirms that higher viscosity suppresses interfacial undulations and promotes more symmetric encapsulation. Experimental studies by Thoroddsen et al. (2005b, 2005a), using ultrafast imaging, showed that the neck curvature during the coalescence of two droplets or bubbles becomes sharper with increasing viscosity, resembling a triangle corner, while at low viscosities, the conjunction region of the neck is more concave with a distinct radius of curvature. At the point of full encapsulation, the dimensionless encapsulation times are $t^*$=4.76 ms and $t^*$=11.4 ms for the lowest and highest droplet viscosities, respectively. Here, to simplify the calculation of encapsulation time, the collision time ($t_{col}$ = 19.68, 20.31, 22.22, and 41.27 ms for cases a–d, respectively) has been subtracted from $t$, such that the normalized time is defined as $t^*$=$t$-$t_{col}$.

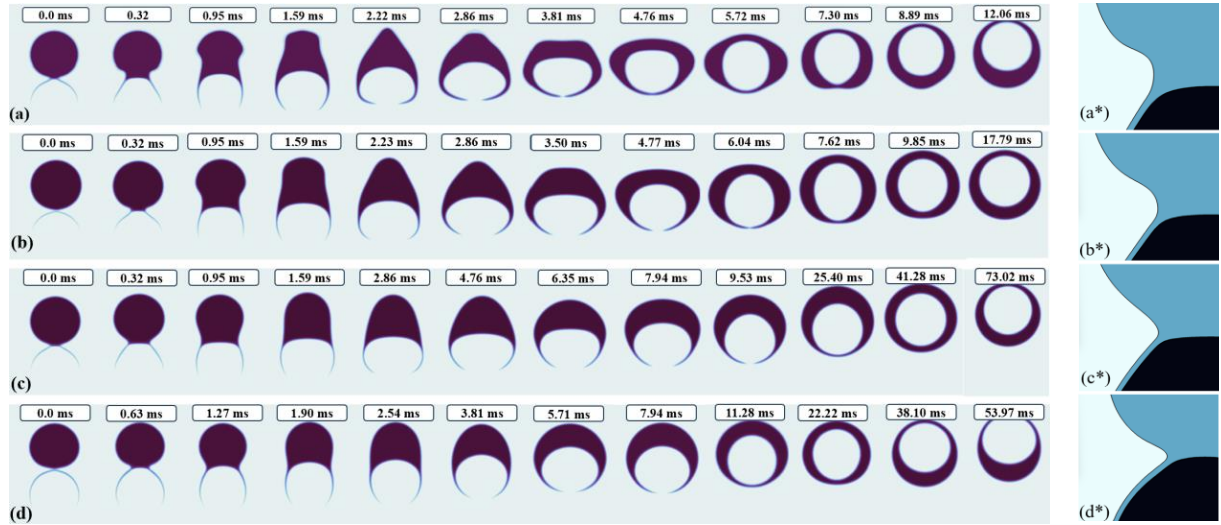

Figure 7. Evolution of the drop's phase-field index $\phi_2$, during bubble encapsulation at different oil viscosities (a-c), and the continuous phase viscosity (d). (-a) $\nu_2 = 0.98 \times 10^{-6}$ m²/s, $\nu_1 = 1.4 \times 10^{-6}$ m²/s , -b) $\nu_2 = 10 \times 10^{-6}$ m²/s, $\nu_1 = 1.4 \times 10^{-6}$ m²/s , -c) $\nu_2 = 56 \times 10^{-6}$ m²/s, $\nu_1 = 1.4 \times 10^{-6}$ m²/s , -d) $\nu_1 = 56 \times 10^{-6}$ m²/s, $\nu_2 = 10 \times 10^{-6}$ m²/s ). For (a) through (d) $Oh_s$=0.037, 0.38, 2.13, and 2.3 respectively. Note that for the latter case viscosity of the continuous phase has been used in the definition of $Oh$. On the right panel, (a*-d*) shows the corresponding close-up of the neck profile at the onset of the encapsulation phenomenon (t*=0.32 ms) at different viscosities.

After full encapsulation of the bubble, its velocity decreases with increasing viscosity, and the rising speed within the droplet also declines. The influence of ambient viscosity is illustrated in Figure 7 (d) for the highest investigated value of the surrounding fluid viscosity ( $\nu_1 = 56 \times 10^{-6}$ m²/s ). The encapsulation dynamics in this scenario resemble those observed for the highest droplet viscosity case (Figure 7 (c)), with one key difference: in the high-viscosity ambient case, the surrounding fluid significantly suppresses bubble deformation. As a result, the bubble maintains a more spherical shape and resists horizontal stretching during the droplet spreading phase ($t^*$=2.54 of row (d)). The maximum vertical deformation ($d_v$) occurs when the bubble is nearly engulfed inside the drop, around $t^*$=5.71 to 7.94 ms. Capillary wave suppression is even more evident in the high-viscosity ambient case than in the high-viscosity droplet case. At $t^*$=3.81 ms in Figure 7 (d) the droplet tip appears noticeably flatter, in contrast to the more conical shapes seen in the upper rows. Additionally, the distance between the upper and lower radial positions of the neck is smaller in this case. This can be visually compared between the neck profiles in d* to c*. According to the relationship for d$R$/d$t$ described in § 2.4.2. , this reduced separation indicates a slower neck growth rate in the presence of a high-viscosity ambient fluid, compared to the case of a high-viscosity droplet.

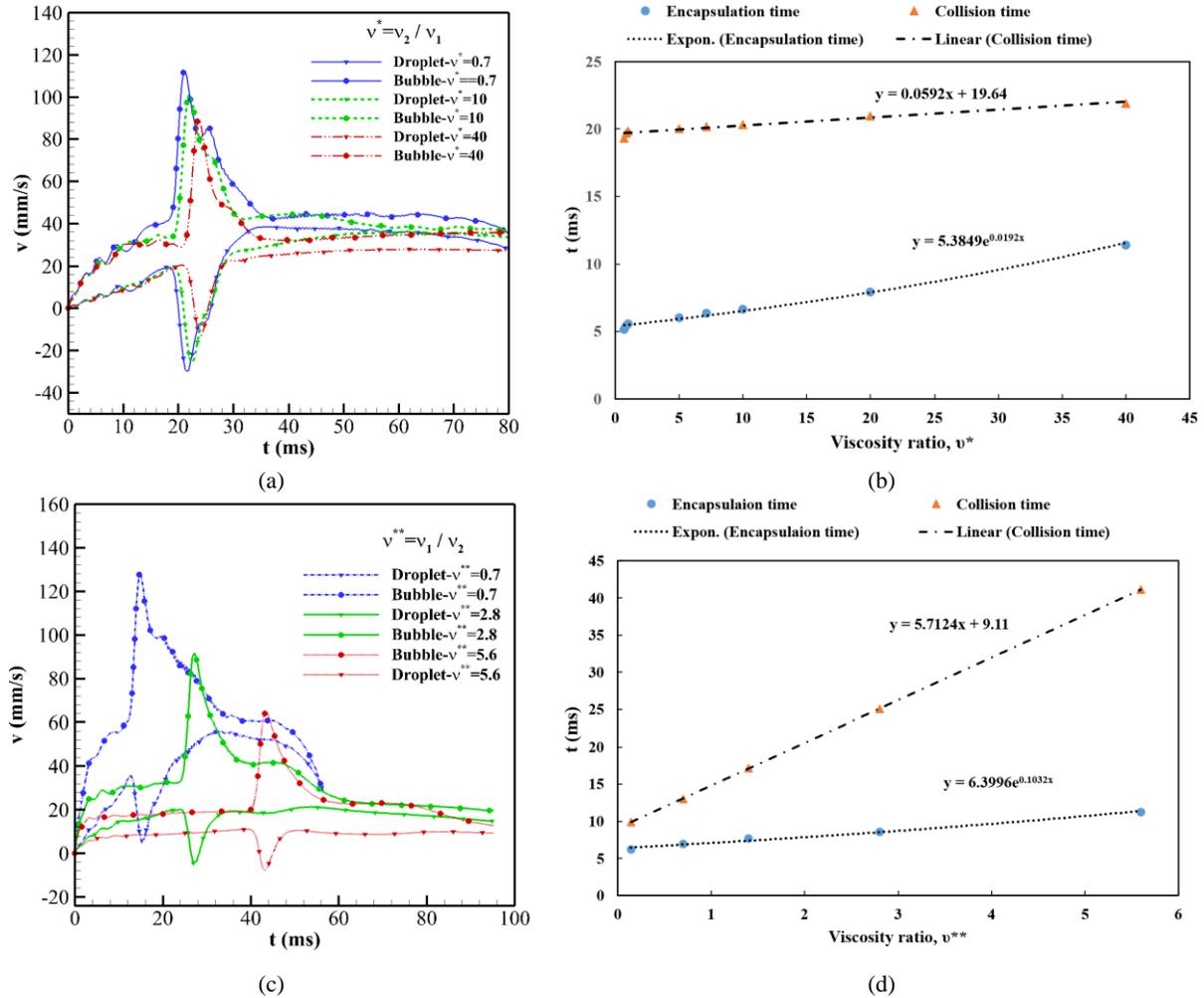

(a)

(b)

(c)

(d)

Figure 8. The effect of viscosity on the droplet and bubble velocity profile during the entire rising and interaction stages. The first row represents; (a) the influence of varying droplet to continuous phase viscosity ($v^*=v_2/v_1$) and the corresponding encapsulation times (b), while the second row shows medium to drop viscosity ($v^{**}=v_1/v_2$) effect on velocity profile (c), and encapsulation times (d). drop and bubble velocity profiles for each viscosity ratio are represented by the same color but different symbols. For case (a) $v_1 = 1.4 \times 10^{-6}$ m$^2$/s , $g=g_0$=9.81 m/s$^2$, and for case (c) $v_2 = 10 \times 10^{-6}$ m$^2$/s , and $g=3.3$ *$g_0$.

Velocity profiles in Figure 8 (a) reveal that increasing the droplet viscosity leads to a reduction in both the maximum bubble penetration and the minimum droplet spreading speed during encapsulation. This trend also holds for the droplet–bubble relative velocity at the moment of impact, which occurs around $t$=20 ms. Specifically, the relative speed drops from over 20 mm/s for the lowest simulated viscosity to approximately 6 mm/s for the highest viscosity case. Another interesting point in Figure 8 (a) is the gradual disappearance of the saddle point near the moment of full bubble encapsulation as the droplet viscosity increases. At high viscosities, this feature transforms into a simple turning point, and one can anticipate that it may vanish entirely at very high viscosities. This behavior stems from the smoother transition of the bubble from a less viscous medium into a more viscous one, which inhibits its ability to quickly reshape and ascend after being completely engulfed. This phenomenon is visually captured in Figure 7, where the detachment of the bubble from the bottom of the droplet after full encapsulation at $t*=7.3$ ms in row (a) and $t*=25.4$ in row (c) illustrates the effect of viscosity on bubble dynamics. Moreover, the velocity profiles in Figure 8 (a) indicate a clear trend that the rising velocity of the oil-coated bubble steadily decreases as droplet viscosity increases.

The droplet–bubble velocity profiles illustrating the effect of carrier phase viscosity are presented in Figure 8 (c). Due to the substantial increase in simulation runtime with higher medium viscosities, this case was conducted under an accelerated gravity of $3.3g_0$ (where $g_0=9.81$) to maintain computational efficiency. A notable reduction in the maximum bubble velocity is observed, indicating that the previously discussed smoothing of the saddle point around the full encapsulation moment, as viscosity increases, is even more pronounced in this scenario. Additionally, the duration of the saddle region becomes longer with increasing medium viscosity. For instance, the period from $t=34$-$50$ ms for $v**=2.8$ contrasts with the much shorter duration near $t=20$ ms for the lower viscosity case in Figure 8 (c). As shown in Figure 8 (b) and (d), both droplet and medium viscosity influence the interaction significantly. An approximate twofold increase in encapsulation time is observed at the highest viscosity ratio in both cases. Interestingly, while both relative collision velocity and the rising speed of the droplet–bubble compound exhibits an inverse relationship with viscosity, increasing droplet viscosity impacts the encapsulation time more substantially than the collision time. Conversely, a higher medium viscosity exerts a stronger effect on prolonging the collision time.

### 3.3. *Interfacial tension effect*

Interfacial tension (IFT) is arguably the most fundamental physical property in a three-phase system, particularly regarding the droplet spreading dynamics. The positivity of the spreading coefficient of a droplet (2.27), is a prerequisite for encapsulation of the bubble. The spreading coefficients also govern the wetting state between the three fluid phases, which can generally result in one of three scenarios: no engulfment, partial engulfment, or total engulfment. In line with the objectives of this study, we focus exclusively on the total engulfment (encapsulation) scenario. To reduce the complexity of the problem and maintain relevance to practical applications, the IFT between the bubble and the continuous phase (water) is kept constant, while the oil-water and oil-bubble IFTs are varied systematically. The simulation setup and the physical parameters used in this section correspond to SS3 of Table 3.

The density contours and the evolution of the droplet's phase-field index are presented in Figure 9 for the highest ($S_o=10.3$ mN/m, $Oh_s=0.143$) and lowest spreading coefficients ($S_o=0.29$ mN/m, $Oh_s=0.85$). It can be seen that surface tension affects both the bubble-droplet shape deformations and the drop's spreading speed. At the moment of initial contact ($t*=0$), a slight shape deformation is already noticeable for the droplet in $S_o=10.3$ case, caused by the lower IFT with surrounding phase. In contrast, the bubble having a higher IFT with drop phase, and also compared to the $S_o=0.29$ case, shows greater resistance against deformation, even during the penetration stage that the bubble experiences its largest velocity in the interaction (note the smoother bubble rear-end for $S_o=10.3$ at $t*=1.27$ ms). This figure also indicate that unlike the base case, the bubble reshaping and full encapsulation stages are not necessarily decoupled at very low spreading coefficients and may overlap. For the case of $S_o=0.29$, the bubble is fully engulfed at around $t^*$=7.30 ms, while in the $S_o$=10.3 mN/m, which is the largest investigated spreading coefficient here, the encapsulation is completed at $t^*$=3.17 ms.

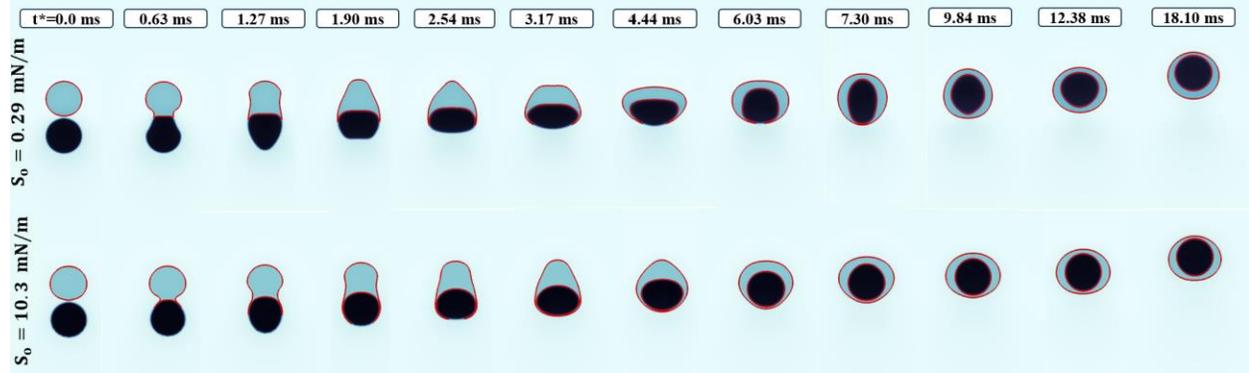

Figure 9. Shape evolution of bubble-droplet interface during encapsulation. The upper row represents a low spreading coefficient case of $S_o$=0.29, $\sigma_{12} = 40.0$ , $\sigma_{23} = 17.64\,\mathrm{mN/m}$ , and the bottom row shows a high spreading coefficient of $S_o$=10.3 mN/m with $\sigma_{12} = 17.64$ , $\sigma_{23} = 29.4\,\mathrm{mN/m}$ . The collision time that has been subtracted from $t$ is 19.36 ms for the former and 22.54 for the latter and $\sigma_{13} = 57.33$ for both cases. For these cases, $Oh_s$= 0.85, and 0.143, respectively. The solid red line is the order parameter of the droplet phase, $\phi_2$ that shows the drop's instantaneous front position during the spreading over the bubble.

Another interesting takeaway in Figure 9 is the high variation of the bubble's aspect ratio ($d_h/d_v$), causing an oscillatory behavior of the bubble all through the penetration to detachment stage (see the magnified area in Figure 10 (b)). For $S_o$=0.29, the bubble reaches its maximum horizontal elongation at $t^*$=3.17 ms, where it loses the kinetic energy gained during the penetration phase. This marks the first local minimum in the velocity profile shown in Figure 10 (b). Following this, surface tension acts to restore the bubble's circular shape, reducing its surface energy. This recovery leads to a regain in kinetic energy and an increase in vertical velocity, initiating the detachment stage. This is reflected as the second local velocity peak, observed between $t^*$=7.30 to 9.84 ms, where the bubble is vertically stretched ($d_h/d_v$<1). This energy exchange cycle occurs again after detachment at $t^*$=12.38 ms and continues a few more times with diminishing intensity as the bubble ascends toward the apex of the drop. In contrast, the bubble in the $S_o$=10.3 case, due to its higher surface tension with the oil drop ( $\sigma_{23} = 29.4\,\mathrm{mN/m}$ compared to $\sigma_{23} = 17.64\,\mathrm{mN/m}$ in the $S_o$=0.29 case) maintains a more spherical shape during both penetration and detachment phases, with $d_h/d_v\approx1$. This is reflected in its smoother velocity profile in Figure 10 (b), which lacks distinct saddle points or oscillations. However, upon closer inspection, a slight velocity gradient can be detected following full encapsulation and during the reshaping stage around $t$=28 ms in Figure 10 (b) (corresponding to $t^*$=4.44-6.03 in Figure 9). The bubble retains this rounded shape even during its ascent within the droplet.

The effect of the droplet phase spreading coefficient ($S_o$) on collision and encapsulation times is illustrated in Figure 10 (c). As discussed earlier, encapsulation time shows an inverse relationship with $S_o$, decreasing as the imbalance in interfacial tensions (IFTs) among the three phases increases. This trend is directly attributed to the increase in capillary speed ($U_c$=$S_o/\mu$), where the viscosities of both the droplet and surrounding medium are constant, and the Bond number remains small ($Bo_{d,b}<1$). This finding aligns with previous studies in microfluidic systems (Lee *et al.* 2016; Zarzar *et al.* 2015; Zhang *et al.* 2015) and other ternary fluid configurations (Farajzadeh *et al.* 2012; Moosai & Dawe 2003; Oliveira *et al.* 1999), highlighting why surfactants are frequently employed to lower IFT, thereby promoting coalescence or controlling the spreading rate of oil droplets in industrial and scientific applications. The collision time on the other hand is directly proportional to the $S_o$ and increases with a gentle slope of 0.1. Theoretical models, such as the Hadamard–Rybczynski relation for spherical fluid particles at small $Re$ numbers (Parkinson *et al.* 2008), and empirical correlations (Bäumler *et al.* 2011) for droplets smaller than 3 mm, show that the drag coefficient acting on a rising particle depends on the $Re$ number, and viscosity of the medium and drop/bubble (considering the internal circulation of the fluid). As revealed by velocity profiles in Figure 10 (a), (b), the IFT has negligible impact on the rising speed of the bubble and droplet at the early stages of the ascend. Therefore, the contributing factor in increased collision time with $S_o$ is primarily the greater droplet deformation resulting from lower IFT with the continuous medium. For instance, Figure 9 shows that at $t*$=0, the aspect ratio of the droplet is $d_h/d_v \approx 1$ for $S_o$=0.29, whereas $d_h/d_v \approx 1.1$ for $S_o$=10.3. This minor deviation from a spherical shape leads to a flattened contact area, which results in lower capillary pressure driving the drainage of interstitial liquid from the thin film region, ultimately prolonging both drainage and collision time.

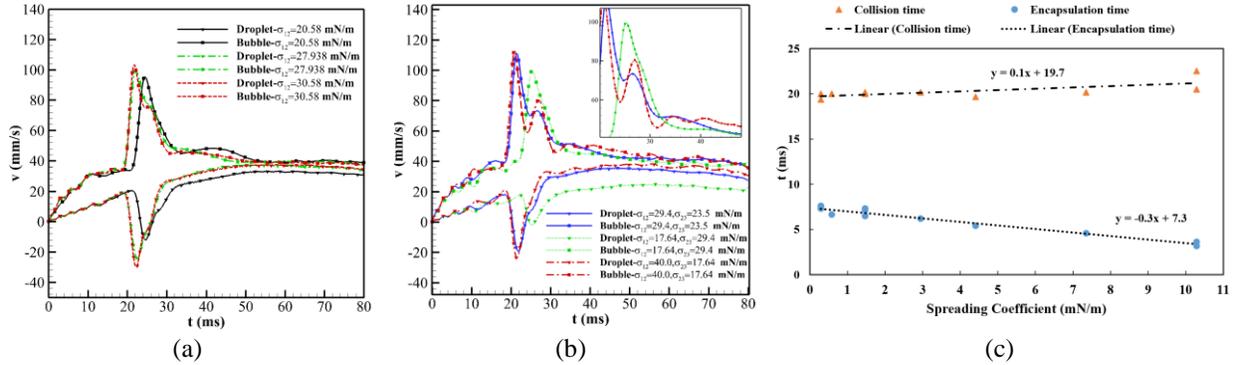

(a)  (b)  (c)

Figure 10. The effect of changing oil drop-medium interfacial tension (a), and simultaneous oil-medium and oil-bubble interfacial tensions (b) on velocity profiles of bubble-droplet. (c) collision and encapsulation time variations with spreading coefficient of oil droplet phase $S_o$. The IFT between the bubble and surrounding fluid is constant for all cases ($\sigma_{13} = 57.33$ mN/m ).

### 3.4. *Effect of droplet density*

In this section, the effect of droplet density on the rise and encapsulation process is examined by varying the density of the oil phase from 700 to 950 kg/m³. This range reflects the typical density of oils used in practical applications and is intentionally kept lower than that of the continuous (aqueous) phase, ensuring that the droplets rise rather than settle. All other physical properties remain consistent with the base case, and the corresponding dimensionless numbers are provided in SS4 of Table 3. The density and phase index contours for the lowest and highest values of $\rho_d$ show no significant differences, closely resembling those observed in the base case (as discussed in §3.1. ). Therefore, these contour plots are omitted here for brevity. However, the velocity profiles in Figure 11 (a) reveal notable trends: as droplet density increases, both the maximum bubble penetration velocity and the minimum droplet spreading speed decrease. Despite these changes in magnitude, the shape and duration of the saddle point during the encapsulation process remain essentially unaffected by variations in droplet density.

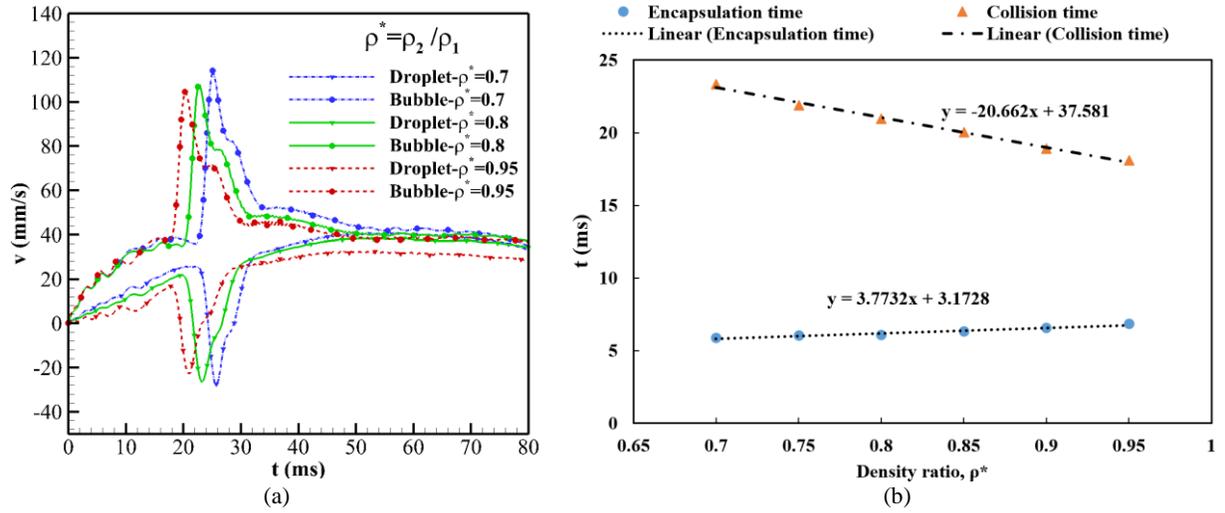

Figure 11. (a) The effect of changing drop density ($\rho_2$) on the vertical velocity profiles of the droplet and bubble with time. (b) The effect of oil drop density on the impact and encapsulation times.

The impact velocity (relative velocity) can be obtained as the difference in the drop and the bubble speed right at the moment of contact. As shown in Figure 11(a), a shorter collision time and higher impact velocity are observed for the case of $\rho^* =0.95$. This is an expected outcome since according to theoretical relations, increasing droplet density (for cases where $\rho_2/\rho_1<1$) leads to reduced buoyancy, thereby slowing the droplet's rise and decreasing the time required to collide with the bubble. Accordingly, collision time decreases as the droplet density approaches that of the continuous medium. The film rupture and the onset of drop spreading occur at around 23.49 ms for $\rho^*=0.7$, compared to 18.73 ms for $\rho^*=0.95$. The lowest impact velocity, around 10 mm/s, is observed at the smallest density ratio $\rho^*=\rho_2/\rho_1=0.7$, while this value nearly doubles to approximately 20 mm/s for the denser droplet case $\rho^*=0.95$. Interestingly, despite the higher impact velocity in denser droplets, encapsulation time does not decrease, as it is inferred from Figure 11 (b). According to (2.36), the neck growth rate is inversely proportional to the droplet density, implying that a smaller density ratio (e.g., $\rho^*=0.7$) facilitates faster neck expansion and thus shortens the encapsulation time, especially since the droplet and bubble sizes are identical. As droplet density increases, encapsulation time increases, showing a positive slope, in contrast to the behavior of collision time. This increase in encapsulation time can also be attributed to enhanced inertial resistance and dynamic pressure, both linked to the droplet's density, which opposes droplet spreading. This finding is in line with the experimental observations of Li et al. (2014) where the authors reported that denser droplets suppress capillary wave formation, thereby hindering satellite formation during bubble passage through a curved liquid-liquid interface.

After investigating the effect of thermophysical properties, the subsequent sections will explore the influence of bubble-droplet diameter and acentric (off-center) interaction on the bubble encapsulation dynamics. The parameter ranges and corresponding dimensionless numbers for these studies are listed in Table 4. To explore the transition from spherical to more deformed regimes (i.e., at higher Bond numbers) and to highlight their differences, simulations in this part are conducted under both standard ($g=g_0$) and enhanced gravity ($g=3.3g_0$). We will refer to the former as the spherical regime, and to the latter as the gravitational regime.

| variable | $R_d$ (mm) | $R_b$ (mm) | $g$ (m²/s²) | $B$ | $Bo_b$ | $Bo_d$ | $Oh_S$ |
|---|---|---|---|---|---|---|---|
| **Size effect, Spherical (SS5)** | 0.4-0.8 | 0.4-0.8 | 9.81 | 0 | 0.11-0.44 | 0.032-0.12 | 0.27-0.38 |
| **Size effect, Gravitational (SS6)** | 0.4-0.8 | 0.4-0.8 | 3.3*9.81 | 0 | 0.36-2.23 | 0.1-0.65 | 0.4-0.45 |
| **Off Center, Spherical (SS7)** | 0.4 | 0.4 | 9.81 | 0-0.8 | 0.032 | 0.11 | 0.38 |
| **Off Center, Gravitational (SS8)** | 0.4 | 0.4 | 3.3*9.81 | 0-0.8 | 0.1 | 0.36 | 0.38 |

Table 4. The properties that are varied from the base case (SS0) to investigate the effect of the droplet-to-bubble size ratio and non-head on collision at both spherical (g=g₀) and deformed (g=3.3g₀) shape regimes. The remaining physical properties are the same as the base case, given in section §2.2. Based on these properties, $Mo_b = 2 \times 10^{-10} - 1.3 \times 10^{-9}$, $Re_d = Re_b = 50.57-302.2$ $Mo_d = 2.2 \times 10^{-11} - 1.48 \times 10^{-9}$ and $Ca_s = 0.51 - 2.44$.

### 3.5. *The effect of bubble-droplet size*

Due to the heterogeneity of the oil drop and air bubble, their size affects the dynamics of the interaction by altering the force balance during the collision, coalescence, or engulfment process. When the size of the fluid particles exceeds the capillary length, they are subject to higher Bond numbers (*Bo*), resulting in greater gravitational effects that promote deformation. This deformation can lead to behavior that differs from that of nearly spherical particles. The simulations carried out to investigate this effect are summarized in SS5 and SS6 of Table 4. These simulations maintain the same physical and geometric properties as the base case in SS1 of Table 3. The initial separation distance (ISD) between the bubble and droplet centers is defined as ISD=1.5*($R_d$+$R_b$). Figure 12 presents density contours and phase-field variable evolution (solid red lines) for interactions between bubbles and droplets of varying size ratios. Rows (a) and (b) illustrate cases where the droplet is smaller than the bubble ($R_d/R_b$<1), while rows (c) and (d) correspond to larger droplets ($R_d/R_b$>1). It is observed that as expected, varying bubble-droplet size changes the collision time since it is directly related to the rising velocity of the bubble and droplet and the collision times increase as the $R_d/R_b$ deviates from 1. Moreover, the increased initial separation distance (which is a function of drop and bubble radius) leads to a slightly increased pre-collision distance traveled by the bubble/droplet (Figure 12 (c)). In contrast, the effect of $R_d/R_b$ on the encapsulation time is more complex to explain and requires further examination of the velocity profiles of both the bubble and the droplet.

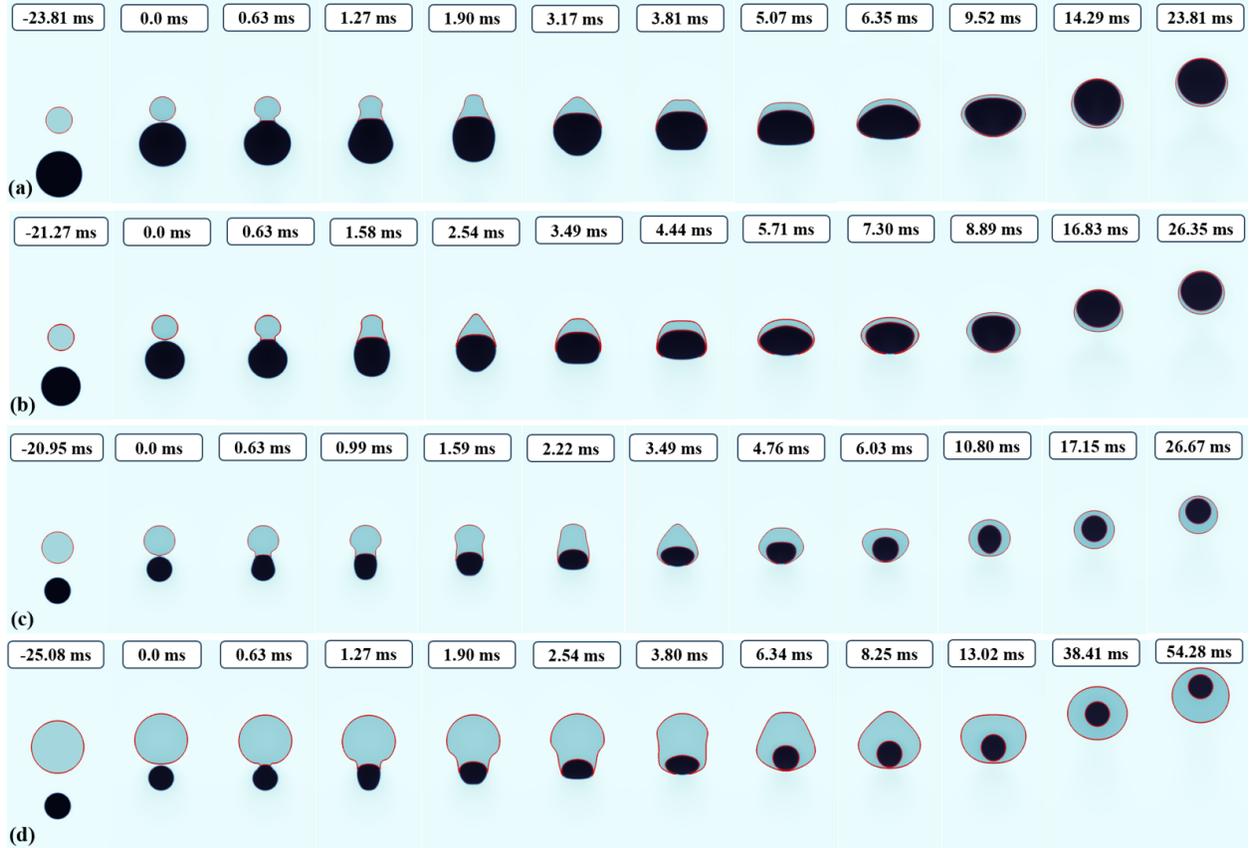

Figure 12. Full-size snapshots of the rise, collision, and bubble encapsulation process at various drop-to-bubble size ratios in spherical regime ($g=g_0$). (a) $R_d/R_b$=0.5, (b) $R_d/R_b$=0.83, (c) $R_d/R_b$=1.2, (d) $R_d/R_b$=2.0. Solid red lines indicate the evolution of the drop interface. The moment when these lines converge at the bottom of the bubble indicates the encapsulation time. Hence, the first time strand of each row represents the collision time that has been subtracted from $t$ for a better comparison of shape evolution and encapsulation stages. The compound $Oh$ and $Bo$ numbers for cases (a) through (d) are $Oh_s$=0.54, 0.41, 0.31, 0.19, $Bo_b$=0.44, 0.157, 0.11, 0.11, and $Bo_d$=0.032, 0.032, 0.046, 0.127, respectively.

### 3.5.1. *Effect of bubble-droplet size on the velocity profile*

For the case of the simultaneous increase in bubble-droplet size, Figure 13 (a) reveals that the contact point of the droplet and bubble shifts towards larger times as their radius increases from 0.4 to 0.7 mm. Additionally, the relative velocity of the collision rises from 0.015 mm/s to 0.028 mm/s. However, the difference between the bubble's maximum velocity and the droplet's minimum velocity during the penetration stage exhibits an inverse relationship with bubble-droplet size. Furthermore, the depth of the saddle point, representing the duration of complete encapsulation, increases with size, extending the encapsulation time. As shown in Figure 13 (b), similar trends are observed under increased gravity of $g$=3.3$g_0$, where the difference between maximum and minimum bubble-drop speed becomes more pronounced, and the saddle region's depth and width expand further. By closely analyzing all simulations and comparing velocity profiles with density distribution contours, it can be deduced that the width of the saddle region is linked to the deformation of the bubble and its deviation from the spherical shape. The closer the bubble is to its spherical shape during the engulfment stage, such as observed behavior in high viscosity droplets or low IFT bubbles, the saddle region is compressed, shrinking towards a point. While, as the bubble deforms into an elliptical shape with an increasing aspect ratio ($d_h/d_v$), the depth and width of the saddle region also increase. This deformation is seen in larger bubbles, where the bubble's internal pressure and surface energy are not high enough to withstand the shape deformations (such as case b Figure 12), or in high $Bo$ number scenarios where inertial forces dominate over surface forces (for example in gravitational regime). It was discussed in the base case simulation that the saddle region corresponds to the stage of complete engulfment of the bubble, and the relative minimum in the bubble velocity profile in this region is equivalent to the duration of the full coverage. After this point, the bubble begins to recover its shape under the effect of surface tension to retain its spherical form (or its equilibrium form depending on the size of the bubble and the physical characteristics of the drop). Consequently, the greater the deformation, the more time is spent on this shape recovery, while the vertical position of the bubble remains constant during this period. Therefore, an increase in the length or depth of the saddle region along the time axis is observed.

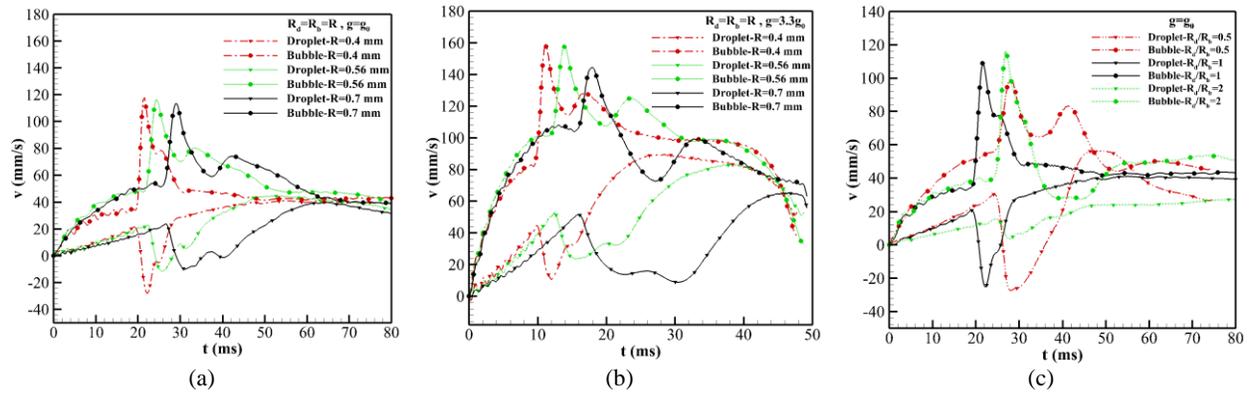

Figure 13. Effect of increasing equal-sized drop and bubble diameter on vertical velocity for (a) $g=g_0$, and (b) $g=3.3g_0$ where $g_0=9.81$ m/s² is the gravity acceleration. (c) variation of velocity with droplet-to-bubble size ratio ($R_d/R_b$) at $g=g_0$.

The velocity profiles for non-equal-sized drops and bubbles are depicted in Figure 13 (c) for different size ratios, ($R_d/R_b$). An interesting point in this figure is the contrasting behavior of drop and bubble in the lowest and highest diameter ratios. The minimum drop velocity (which means the highest downward velocity of the drop) is recorded for $R_d/R_b$=0.5, while the largest bubble velocity (bubbles penetration speed) corresponds to $R_d/R_b$=2. Therefore, for droplets larger than the bubble (for example $R_d$=0.8, $R_b$=0.4 mm), the less impact it receives from the interaction force exerted by the bubble, and the drop's released surface energy mostly translates to the bubble kinetic energy, since the bubble is small and has high internal pressure. In contrast, the smaller the droplet ($R_d$=0.4, $R_b$=0.8 mm) has more surface energy to release, which accelerates its spreading speed, but a large portion of the released energy is absorbed by the bubble and is transformed into its shape deformation. The following explanation may further clarify this phenomenon.

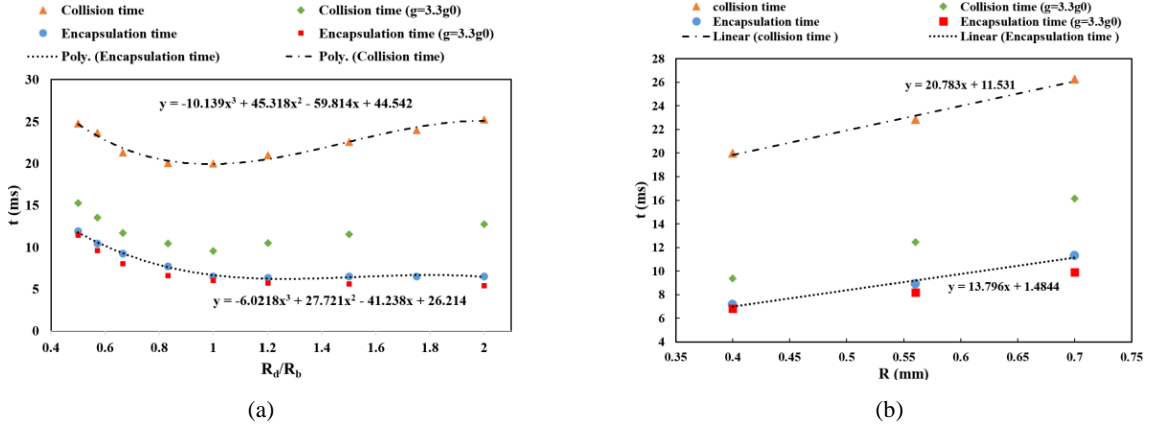

<div align="center">(a)                                           (b)</div>

Figure 14. The influence of drop and bubble size on the collision and bubble encapsulation times. (a) increasing $R_d/R_b$ (unequal sized drop and bubble), (b) increasing bubble and drop radius at constant $R_d/R_b$=1 (equal sized interaction).

### 3.5.2. *Capillary pressure and its influence on the flow direction*

From the Young-Laplace law, the capillary pressure inside a drop or bubble is inversely related to its size ($\Delta p = 2\sigma/R$). Upon contact, a vertical pressure gradient arises due to differences in capillary pressure and the high curvature of the neck at the onset of liquid bridge growth. When the bubble is larger than the drop (for example in $R_d/R_b$=0.5 case where $\Delta p_d = 150\,\mathrm{Pa}$ and $\Delta p_b = 140\,\mathrm{Pa}$ ), the pressure inside the drop is higher than the bubble and the fluid flow is from the drop side to the bubble side. Conversely, for $R_d/R_b$=2, the bubble's higher pressure ( $\Delta p_d = 75\,\mathrm{Pa}$ and $\Delta p_b = 280\,\mathrm{Pa}$ ) drives fluid flow from the bubble to the drop. Based on the capillary pressure difference in the drop and bubble in the previous paragraph, we have:

$$\Delta p_b - \Delta p_d = \frac{2\sigma_{13}}{R_b} - \frac{2\sigma_{12}}{R_d} = \frac{2\sigma_{13}}{R_b}\left(1 - \frac{(\dfrac{\sigma_{12}}{\sigma_{13}})}{(\dfrac{R_d}{R_b})}\right) \tag{3.1}$$

Therefore, taking into account that $\sigma_{12}/\sigma_{13}$ is constant, for $R_d/R_b \gg 1$ the pressure difference that drives the flow approaches towards the capillary pressure of the bubble. In other words, with the increase in the radius of the curvature of the droplet, the capillary pressure inside the drop decreases drastically. This leads to contrasting interactions: in the former case, the bubble pulls the drop toward itself, while in the latter, the drop engulfs the bubble.

### 3.5.3. *Encapsulation time trends*

The effect of the capillary pressure manifests in the calculated encapsulation times. In Figure 14 (a) it can be seen that the duration of coverage for $R_d/R_b$<1 increases with the reduction of the size ratio. The larger bubble surface requires more time for the drop to fully cover it. A similar behavior was observed in coalescence studies of unequal-sized droplets by Deka et al. (2019), where it was indicated that the capillary waves have to travel a longer distance to reach the bottom of the father drop while being constantly dampened by viscous effects. This effect can be observed by comparing cases (a) at $t$*=3.17 ms and (d) at $t$*=1.27 ms of Figure 12, where capillary waves reach and merge at the base of the bubble by forming a triangular tip. These findings align with experimental observations of gasoline droplet-air bubble interactions in §2.4.2. , where encapsulation time was longer for $R_d/R_b < 1$ despite faster neck growth rates. For $R_d/R_b > 1$, the encapsulation time shows negligible variation with the increase in $R_d$ and remains nearly constant as $R_d/R_b \rightarrow \infty$. This is because, as the droplet size increases relative to the bubble, the interface curvature flattens, reducing capillary pressure inside the drop. Consequently, the system resembles a fixed-size bubble penetrating a liquid-liquid interface. Figure 13 (c) further confirms this by showing an increased bubble velocity for $R_d/R_b = 2$, while the droplet exhibits minimal speed reduction in the spreading phase. When the diameters of equal-sized bubbles and droplets increase, not only the bubble surface to be covered increases but also the droplet capillary pressure that drives the spreading decrease, thus resulting in higher encapsulation times (Figure 14 (b)).

### 3.5.4. *Effect of gravitational acceleration on the interaction*

Figure 14 also depicts the influence of increased gravity ($g = 3.3g_0$) on collision and encapsulation times. Higher gravitational acceleration significantly shortens collision time by enhancing buoyancy forces and impact velocity. From velocity profiles of Figure 13 (b), it can be observed that for example the 0.7 mm bubble's velocity at the moment of impact is raised to around $v_b = 105$ mm/s, compared to $v_b = 50$ mm/s for the base $g_0 = 9.81$ m/s² case. This shows that the rising velocity of the bubble and the droplet, as well as their impact velocity (the relative velocity of the bubble and droplet at the moment of impact) almost doubled at $g = 3.3g_0$. However, gravitational acceleration has a much smaller effect on encapsulation time, reducing it by only ~1 ms. This trend holds for equal-sized bubble-droplet pairs, with gravitational effects becoming more pronounced at larger sizes as demonstrated in Figure 14 (b). Aside from above mentioned differences, the general trends of velocity profiles, as well as the collision and encapsulation times are very similar in both gravitational states. To further explore the effect of collision velocity on encapsulation time, another simulation series has been performed by varying the initial separation distance (ISD) from $2.5R$ to $5R$ (see supplementary materials 4). Results indicate that encapsulation time reduces from 7.1 ms at ISD=$2.5R$ (collision velocity = 0.36 mm/s) to 5.9 ms at ISD = $5R$ (collision velocity = 52 mm/s). This suggests that, in the studied range and near-spherical regimes, encapsulation time is more dependent on physical properties like spreading coefficient and viscosity rather than collision velocity. Although no systematic studies on the relationship between encapsulation time and droplet-bubble size exist, prior research on the coalescence of identical fluid particles (Dudek *et al.* 2020; Kamp *et al.* 2017) confirms that coalescence time increases with droplet or bubble size, aligning with our findings.

### 3.5.5. *Equilibrium shapes of the bubble-droplet compound*

Figure 15 highlights the influence of increased gravitational acceleration on the equilibrium shape of the coated bubble during the compound rising stage. Compared to the fixed-$g$ simulations ($g=g_0$, SS5 in Table 4), the larger deformations of the compound in accelerated gravity ($g=3.3g_0$ , corresponding to SS6 of Table 4) are perfectly recognizable in Figure 15 (a). As illustrated in Figure 15 (b), the SS5 cases, characterized by $Bo_b < 0.44$ and $Bo_d < 0.12$, maintain aspect ratios ($d_h/d_v$) below 1.1 across all volume fractions ($\alpha_0$) asymptotically approaching unity as $\alpha_0$ increases. In contrast, the SS6 series, with $Bo_b > 0.36$, begins at an aspect ratio of $d_h/d_v = 1.3$ for $\alpha_0 = 0.11$ and similarly trends toward unity at $\alpha_0 = 0.89$. Notably, the bottom row of Figure 15 (a) reveals that at high $\alpha_0$ values, the bubble undergoes greater deformation than the compound, adopting an oblate ellipsoidal shape with an enlarged horizontal axis. Based on these observations, the SS5 simulations may be reasonably classified as representative of the spherical regime, while SS6 captures a distinctly deformed regime. The transition between these regimes appears to occur near $Bo_b \approx 0.4$ and $Bo_d \approx 0.1$. Interestingly, the trend in aspect ratio under increased gravity closely resembles experimental values reported by Ji et al. (2022) for $D_b \approx 4$ mm coated with 10 cSt silicon oil at various oil fractions.

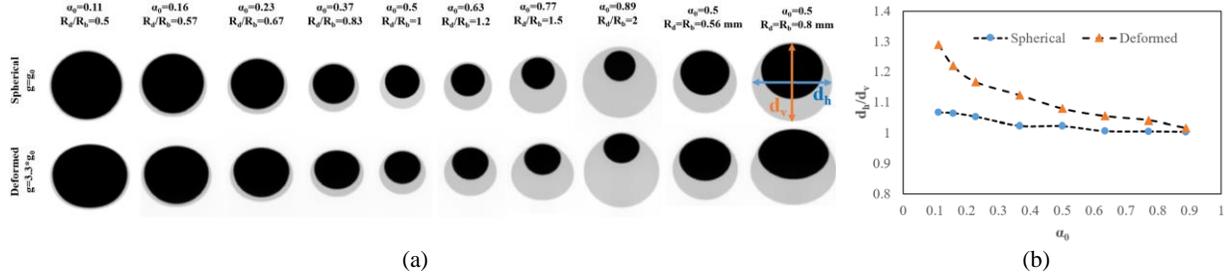

(a)                                                                                                                          (b)

Figure 15. (a) The equilibrium shapes of the bubble-droplet compound after encapsulation in spherical (first row: $Bo_b < 0.44$ and $Bo_d < 0.12$) and deformed/gravitational (second row; $Bo_b > 0.36$ and $Bo_d > 0.1$) regimes. The time instance in the former is $t=63.5$ ms, while $t=47.6$ ms in the latter. (b) the horizontal to vertical aspect ratio of the compound at different oil volume fractions $\alpha_0 = V_{oil}/(V_{oil}+V_{bubble})$. The equivalent size ratios are also included in the legend of the figure. The physical properties of these cases correspond to SS5 and SS6 in Table 4.

### 3.6. *Off-center (non-head-on) interaction*

In real systems, most particle collisions, including those involving bubbles and droplets, are not perfectly centered and often occur at an angle. This section investigates the impact of such off-center interactions on collision and encapsulation dynamics. Simulations were conducted under two gravitational conditions: $g=g_0$ (SS7) and $g=3.3g_0$ (SS8), where the impact parameter B (defined in Figure 1 (a)) was varied in the range of 0 to 0.8. The simulation properties for these tests are summarized in Table 4, with physical properties and dimensionless numbers matching the base case (SS0 of Table 3).

### 3.6.1. *Shape evolution and velocity profiles for varying impact parameters*

Figure 16 compares the simulated evolution of bubble-droplet shapes during an eccentric encapsulation process with experimental images, at a relatively similar setting. The shape differences between the bubble-droplet compound and those formed during a head-on collision are minimal, with only slight asymmetry observed in the bubble's position within the droplet. For example, at $t=4.127$ ms, the bubble appears closer to the right wall of the drop. This minor asymmetry is a result of the interaction occurring in the spherical regime, where both the bubble and droplet sizes are below the capillary length ($L_{cb}=2.42$ mm, $L_{cd}=3.47$ mm), thus the gravitational effects are negligible. In contrast, the experimental results exhibit a more noticeable asymmetry, as the bubble and droplet sizes are larger and closer to the deformed regime, where gravitational forces become more influential. The discrepancy in encapsulation speed between numerical and experimental results can be attributed to the differences in the nature of the interactions: in the numerical simulations, the interaction is dynamic, with relative velocity initiating film drainage and droplet spreading. In the experimental setup, the process is quasi-static, with spontaneous drainage and negligible initial velocity.

Additionally, in the experiments, droplets are formed at the tip of a capillary, leading to lower surface energy compared to free-floating droplets. This difference is particularly noticeable for this specific case, as the liquid bridge formed after film rupture is closer to the drop's generation nozzle. In comparison, for interactions described in §2.4.2. , spreading initiated at the furthest point from the nozzle, where the droplet had a larger radius of curvature and higher interfacial energy. Despite these differences, the simulation results generally align with experimental observations in terms of shape evolution and the stages of encapsulation.

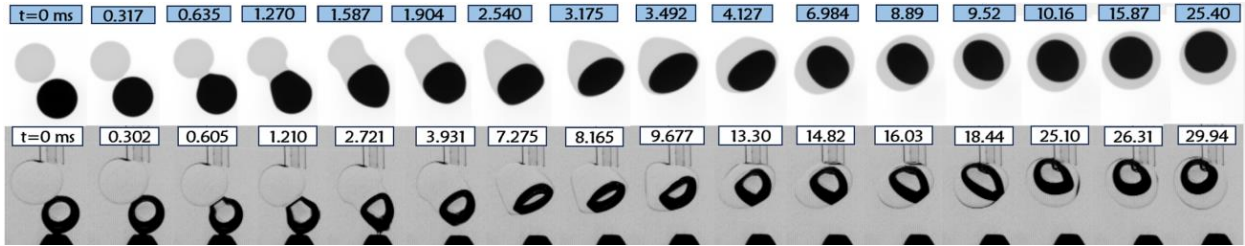

Figure 16. Off-center encapsulation of the bubble by the oil droplet. The upper row represents numerical snapshots for $B$=0.4, corresponding to SS7 of Table 4. The second row represents experimental sequences of quasi-static interaction between an air bubble with $D_b$=1.72 mm and 10 cSt silicon oil droplet with $D_d$=2.2 mm. The calculated collision parameter for the experiments is $B_{exp}$=0.45, and its physical properties are given in Table 1.

Density contours and velocity vectors for the bubble-droplet interaction at $g$=3.3$g_0$, and collision parameters of $B$=0.6 and $B$=0.8 are displayed in Figure 17 (a). It can be seen that gravitational effects come into play in this case, leading to higher asymmetry and larger deformations in bubble-droplet shapes. For $B$=0.6, the contact is formed at $t$=13.01 ms, with the bubble's velocity primarily directed upwards. This results in a more intense concentration of velocity vectors on the upper side of the aggregate during the encapsulation phase ($t$=13.65-17.46 ms). Over time, surface tension gradually compensates for this asymmetry, and the flow field becomes more uniform after $t$=19.05 ms, however, the hydrodynamic force applied by the bubble at the moment of the impact causes the compound to deviate from its path and leaning to the left, a behavior that was less pronounced in the spherical regime or smaller impact parameters. For the case of $B$=0.8 (second row in Figure 17 (a)), the bubble exerts higher shear forces and a greater hydrodynamic impact on the droplet, causing it to deviate further from its original path and move toward the left wall. Interestingly drop's vertical position remains unchanged during the close encounter ($t$=13.01-$t$=19.05 ms). Yet, induced shearing force by the bubble slightly stretches the droplet, and once the bubble passes by, it creates a downward velocity gradient that pushes the droplet down at $t$=25.87 ms.

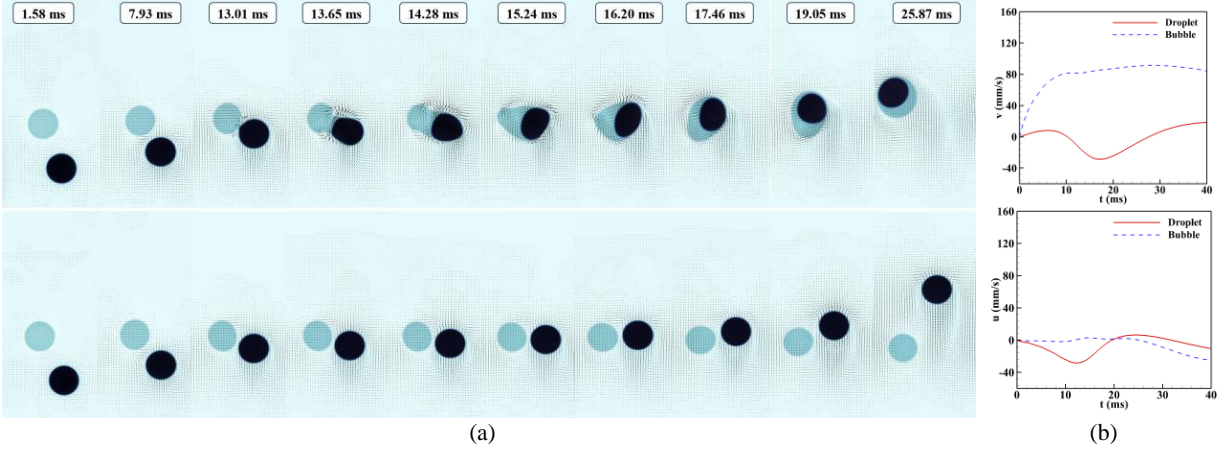

(a)

Figure 17. (a) Sequences of rising and interaction between the bubble and the droplet in the collision parameter of $B$=0.6 (top row) and $B$=0.8 (bottom row) at $g$=3.3$g_0$. In contrast to the spherical regime ($g$=$g_0$), in this case, no collision and engulfment were observed for $B$>=0.75. (b) vertical (top) and horizontal (bottom) velocity of the bubble and droplet in unsuccessful interaction at $B$=0.8.

Recent experimental investigation by Federle et al. (2024) shows that the bubble-induced flow field can generate a 2-3 times increase in droplet's rising, with the largest velocity increases occurring at smaller collision angles and droplet-to-bubble diameter ratios. They approximated the flow around the bubble as an Oseen-type solution, while the droplet response was modeled using the Maxey-Riley equation. The simulated bubble and droplet's velocity for the non-contact interaction at $B$=0.8 is depicted in Figure 17 (b). In contrast, the current study shows a more modest increase in droplet velocity after the bubble-induced shock. This is likely due to the fact that both the bubble and droplet are the same size here, and the collision angle is relatively large. Furthermore, Federle et al.'s bubbles were between 2–3 mm in size, which provided more momentum and a larger wake, leading to a greater impact on the droplet's motion. Figure 17 (b) also reveals that the droplet experiences a sudden acceleration in the negative direction of both the horizontal and vertical axis. This is where the bubble's rise velocity stays nearly constant after reaching a terminal velocity of 80 mm/s at around $t$=10 ms. After the interaction, both the bubble and the droplet exhibit a slight deviation towards the left boundary, with their horizontal velocities becoming negative starting around $t$ = 20 ms. This is due to the transfer of momentum from the vertical to the horizontal direction, a result of the hydrodynamic interaction. This path deviation is even more pronounced in the $B$ = 0.6 case, where the compound eventually exits the simulation domain at $t$ = 25.87 ms following successful collision and encapsulation (top row of Figure 17 (a)). This phenomenon can be further understood through the profiles of the horizontal velocity of the drop and bubble, as shown in Figure 18 (a) and Figure 18 (b). The key observation here is that increasing the collision parameter $B$ leads to the conversion of vertical velocity ($v$) into horizontal velocity ($u$) during the encapsulation phase. This can be confirmed by comparing $v$ and $u$ at the maximum and minimum drop-bubble speeds for $B$ = 0 and $B$ = 0.6. As B increases, larger portions of the momentum are transferred from the vertical to the horizontal direction. Moreover, the saddle region also disappears from $v$ and reappears in horizontal velocity $u$ with greater width and depth. This momentum shift, combined with the off-center collision, justifies the observed deviation of the aggregate after encapsulation.

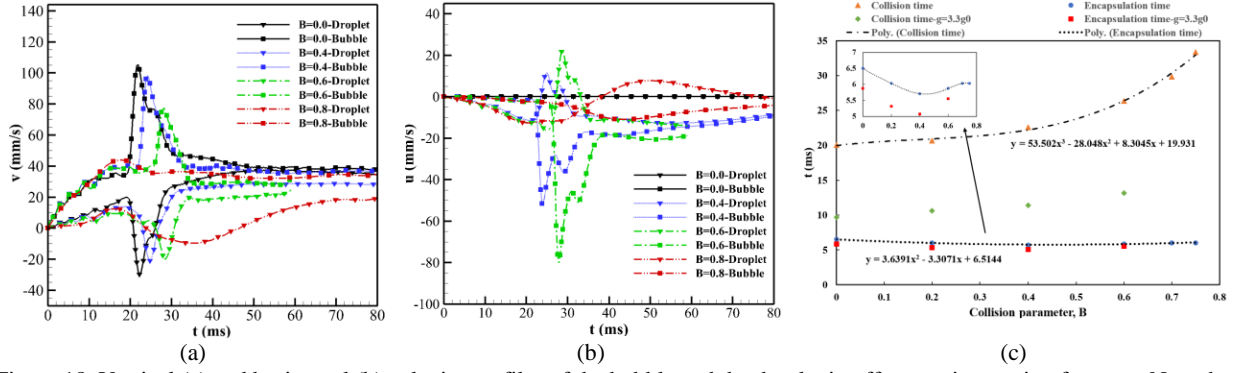

Figure 18. Vertical (a) and horizontal (b) velocity profiles of the bubble and the droplet in off-center interaction for $g=g_0$. Note that in the horizontal velocity u, the lower profiles belong to the bubble, where its direction at the penetration stage is towards the left (negative axis values). Moreover, no physical contact occurs at $B=0.8$. (c) The effect of the impact parameter on the collision and encapsulation times in spherical ($g=g_0$) and gravitational ($g=3.3g_0$). The small image within the figure signifies the encapsulation time in a zoomed-in mode. Physical properties are the same as the base case (SS0) of Table 3, and other parameters SS7 and SS8 of Table 4.

### 3.6.2. *The influence of eccentric interaction on collision and encapsulation times*

The characteristic collision and encapsulation times are shown in Figure 18 (c) for gravity accelerations of $g=g_0$ and $g=3.3g_0$. In both cases, collision time increases with $B$, particularly at higher values of $B$ ($B > 0.5$). This is due to the hydrodynamic forces and shear effects, which hinder film drainage and contact formation compared to head-on collisions. For encapsulation time, the trend follows an initial decrease to a minimum (around 5.7 ms for $g = g_0$, and 5.1 ms for $g = 3.3g_0$) at $B \approx 0.5$, corresponding to a collision angle of about 45°. After this minimum, encapsulation time increases again as $B$ continues to rise. This non-monotonic behavior likely arises because of momentum transfer from vertical to horizontal directions, as discussed in the previous section. At $B\approx0.5$, the bubble and droplet momentum in the horizontal and vertical directions are approximately equal. This indicates that the optimal impact angle for bubble penetration and droplet spreading is near 45°. For example, at $B = 0$, the maximum bubble velocity is $u \approx 0$ and $v = 100$ mm/s, while at $B = 0.6$, the bubble's horizontal velocity is $u = -80$ mm/s and vertical velocity is $v = 75$ mm/s (Figure 18 (a)-(b)). This difference is partly because when the bubble is directly behind the droplet, its rising velocity is suppressed by the droplet's presence, while the droplet's velocity is slightly enhanced as the bubble pushes it upward. For eccentric interactions (larger $B$), the bubble approaches the droplet from the side, resulting in a slightly higher velocity at impact. At the same time, the droplet's velocity is lower than the head-on case, leading to a greater relative impact velocity during off-center collisions. This can be inferred from Figure 18 (a) by comparing bubble and droplet velocities for $B=0$ and $B=0.4$, 0.6 between $t=20$-26 ms. This is also consistent with the findings from Federle et al. (2024), which suggested that the maximum increase in droplet velocity occurs at the smallest collision angles.

## 4. Unifying the results through scale analysis

After presenting the parametric study, in this section, we will perform a force balance and scale analysis to provide theoretical insight into the modeling of bubble-droplet interaction. Regarding the collision time, the numerical results appear to be more definite and predictable, and all the trends of the impact time can be justified based on the initial distance, size, and relative speed of the bubble-drop. The problem has also been studied to some extent in the literature (Federle *et al.* 2024; Kordalis *et al.* 2023; Wang *et al.* 2024a, 2024b). On the other hand, the encapsulation time depends on both physical and hydrodynamic properties, making its behavior more complex to analyze. The compound bubble-droplet morphology at the onset of droplet spreading is depicted in Figure 19. The forces driving or assisting encapsulation (spreading of the droplet) are interfacial tension $F_\sigma \sim [(\sigma_{og} + \sigma_{om}) - \sigma_{mg}]R_b \sim S_o R_b$, and buoyancy of the bubble $F_{Bb} \sim (\rho_m - \rho_g)gR_b^3$, while the resisting forces comprise viscous forces (including the viscosity of both the medium and droplet phases) $F_{Vis} = F_{Vo} + F_{Vm} \sim (\mu_m + \mu_o)U_{sp}R_b$, inertial forces $F_{Inert} \sim \rho_m U_{sp}^2 R_b$ acting along the bubble's perimeter (proportional to $R_b$), and buoyancy of the droplet $F_{Bd} \sim (\rho_m - \rho_o)gR_d^3$. Given the complexity of the phenomena, and considering that the droplet's density is of the same order of magnitude as the medium (three orders higher than the bubble), we neglect the fact that, especially in the later stages of encapsulation, the bubble's buoyancy consists of two components: the portion still in contact with the medium and the volume already engulfed within the droplet.

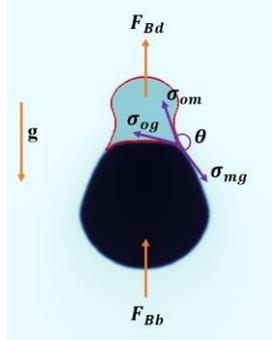

Figure 19. Bubble-droplet configuration and the acting forces during the encapsulation process.

where $U_{sp}$ is the spreading velocity of the droplet. By ignoring $F_{Bd}$ compared to the relatively bigger $F_{Bb}$, and noting that $\rho_m \gg \rho_g$, the force balance between driving and resistance effects can be expressed as

$$S_o R_b + \rho_m gR_b^3 \sim (\mu_m + \mu_o)U_{sp}R_b + \rho_m U_{sp}^2 R_b \qquad (4.1)$$

For the spherical regime, i.e. the tests in which $R_b = R_d$ and only the physical properties were varied, $D_{d,b} < \lambda_{cd,cb}$ (where $\lambda_c = \sqrt{\sigma/\rho g}$ is the capillary length), thus gravitational effects on the droplet's spreading can be neglected. By dropping buoyancy force, and also for simplicity inertial forces (4.1) reduces to

$$S_o R_b \sim (\mu_m + \mu_o)U_{sp}R_b \qquad (4.2)$$

$$\Rightarrow U_{sp} \sim \frac{S_o}{(\mu_m + \mu_o)} = U_{VC} \qquad (4.3)$$

Here, $U_{VC}$ is the viscous-capillary velocity scale, and the corresponding time scale can be expressed as $\tau_{VC} = (\mu_m + \mu_o)R_b/S_o$. The calculated encapsulation times versus $U_{VC}$ are shown in Figure 20 (a), where the relationship can be approximated as a $t_{enc} \sim 4.9\ U_{VC}^{-0.155}$. From another point of view, this implies that $t_{enc}$ increases as $\tau_{VC}$ rises. In §2.4.2. we observed that for relatively low-viscosity gasoline droplets, the neck growth rate, indicating the droplet's spreading speed, scales with the capillary-inertial time scale ($\tau_{CI}$). In an interesting study, Shen *et al.* (Shen *et al.* 2018) investigated the reverse case, where a droplet is encapsulated inside a bubble, and identified a critical $Oh = 0.052$ as the transition point between viscous-capillary and capillary-inertial dominance. Similarly, (Dong *et al.* 2017) suggested $Oh = 0.01$ as the threshold between inertia-capillary and viscous regimes for droplet-pool coalescence in a surfactant-free system. In the numerical tests presented in Table 3, most cases exhibit $Oh_s > 0.1$, with only three tests showing a small $Oh$ number ($Oh_s < 0.1$). Alternatively, if we apply the original definition of $Oh_d$ (using $\sigma_{om}$ instead of $S_o$), the $Oh$ number for the three lowest viscosity cases falls within $Oh_d = 0.0083$–$0.012$, while for the remaining test cases, it ranges from $Oh_d = 0.06$–$0.47$. This justifies the use of the viscous-capillary time scale while neglecting inertial terms. However, the effects of inertia and capillarity on encapsulation time can still be inferred from the $Oh$-$t_{enc}$ relationship as depicted in Figure 20 (b). It can be observed that as $Oh$ increases and viscous effects dominate over inertia-capillary forces, the time required for the droplet to fully engulf the bubble increases.

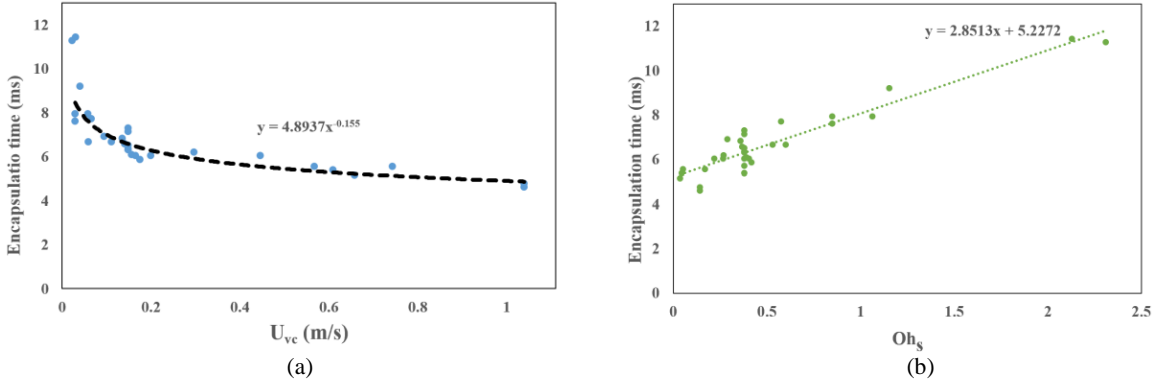

(a)            (b)

Figure 20. Comparison and scaling of the calculated encapsulation times with the viscous-capillary velocity (a) and $Oh$ number (b) for the simulation cases where the drop-bubble size is constant $R_d=R_b=0.4$ mm and the quantities of density, viscosity, and surface tension have been varied. Note that for this set of simulations, $Bo_b=0.11$ and $Bo_d=0.01$-$0.064$, thus the gravitational effects are negligible.

Now, for the simulations where the acceleration of gravity and the size of the droplet-bubble and accordingly changes in the bond number are noticeable, the bubble-droplet interaction is no longer in the spherical regime and is under the relative dominance of gravity, inertia, and capillary effects. Thus, for this deformed regime, introducing $\tau_{IC} = \sqrt{\dfrac{\rho_m R_b^3}{S_o}}$, $Bo_s = \dfrac{\rho_m g R_b^2}{S_o}$, and taking $\mu_e = \mu_o + \mu_m$ and also $T = 1/t_{sp}$, (4.1) can be recast as

$$U = \frac{L}{T} \Rightarrow t_{sp} \sim R_b/U_{sp} \tag{4.4}$$

$$S_o\left(\rho_m g R_b^2 \middle/ S_o + 1\right) \sim \mu_e R_b^2 T + \rho_m R_b^3 T^2 \tag{4.5}$$

$$\tau_{IC}T^{2} + \left(\frac{\mu_e R_b^{2}}{S_o}\right)T - (1 + Bo_s) = 0 \qquad (4.6)$$

Solving the above equation for $T$ we would get

$$T = -\frac{\mu_e}{2\rho_m R_b} \pm \sqrt{\frac{1}{\tau_{IC}^{2}}(1 + Bo_s + \frac{R_b^{2}}{4}Oh_s^{2})} \qquad (4.7)$$

And now for simplicity, if we drop viscous effects, we arrive to

$$\frac{1}{t_{sp}} = T \sim \sqrt{\frac{1}{\tau_{IC}^{2}}(1 + Bo_s)} \qquad (4.8)$$

$$\Rightarrow t_{sp} \sim \frac{\tau_{IC}}{\sqrt{1 + Bo_s}} = \tau_{CG} \qquad (4.9)$$

Here $\tau_{CG}$ is the capillary-gravitational time scale, which is very similar to the proposed relationship by (Chen *et al.* 2006) for the coalescence of a droplet with a miscible liquid pool. Notably, in equation (4.9), $t_{sp}$ approaches the capillary-inertial time scale, $\tau_{CI}$ as the bond number tends to zero ($Bo_s \rightarrow 0$). For the partial coalescence of a drop with a liquid-air interface, it has been suggested that $Bo = 0.1$ is the boundary between gravitational and capillary-inertial regimes (Dong *et al.* 2017). As shown in Figure 21 (a), encapsulation times in the deformed regimes generally follow the capillary-gravitational time scale. However, due to numerical limitations, the bubble-medium surface tension was set to 57.6 mN/m, resulting in $S_o = 1.47$ mN/m, an order of magnitude smaller than the actual surface tension. Consequently, this led to larger $Bo_s$ values and hence a little bit of scatter in data points. Therefore, using $Bo_s$ would be more meaningful if the exact real-world surface tension values were applied. As depicted in the inset of Figure 21 (a), using the original definition of $Bo$, i.e. (2.31) in (4.9) improves the estimation, and the data points are more concentrated around the trendline. However, the trendline equation and its slope are very similar in both cases.

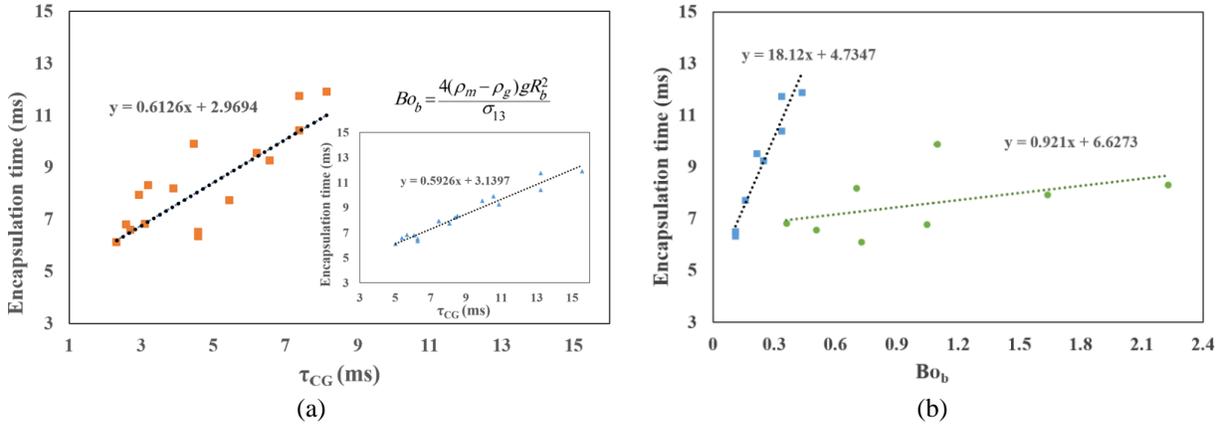

Figure 21. (a) Calculated encapsulation times versus capillary-gravitational time scuale for simulation cases where the droplet-bubble size or gravity acceleration have been changed and the physical properties were kept constant. The smaller figure shows the encapsulation time with $\tau_{CG}$ that has been calculated using $Bo_b$ instead of $Bo_s$. For this set of simulations, $Oh_d < 0.1$, $Oh_s < 0.38$, which justifies ignoring viscous effects. (b) Encapsulation time versus $Bo_b$. Squared blue points represents the case of SS5 of Table 4, while the green circles belong to SS6 tests.

Figure 21(b) demonstrates that the encapsulation time ($t_{enc}$) generally increases with $Bo_b$. Two sets of simulations, SS5 and SS6 of Table 4, have been distinguished here. In SS5, where $R_d/R_b$ is varied at constant gravitational acceleration ($g=g_0$), the encapsulation time exhibits a sharp increase with $Bo_b$ and minimal scatter. In contrast, the SS6 series, which simultaneously varies both $g$ and bubble-droplet size, yields a more gradual increase in $t_{enc}$ accompanied by greater variability. These deviations arise from contrasting physical effects. As discussed in §3.5. , increasing $Bo_b$ through gravitational acceleration ($g$) for a fixed bubble-droplet size tends to reduce $t_{enc}$, owing to the enhanced buoyancy-driven motion. In contrast, an increase in $Bo_b$ by enlarging the bubble radius ($R_b$) at constant gravity leads to longer encapsulation times, since a larger interfacial area must be covered during the engulfment process. Consequently, in the fixed-$g$ simulations, the rise in $t_{enc}$ is primarily driven by increasing $R_b$, yielding a steeper $Bo_b$ dependence. In the variable-$g$ cases, however, the larger $R_b$ effect is partially offset by stronger buoyancy forces, resulting in a milder slope. A similar competing effect was recently encountered by Cuttle et al. (2021), investigating the engulfment of micro-and macrodroplets in a wetting oil pool. The authors report that at high $Bo$, macrodrops cause pronounced surface deformations in the oil layer, promoting faster engulfment. However, at low $Bo$, gravity plays a minimal role, and increasing droplet size merely expands the interfacial area to be covered during spreading. Lastly, it should also be noted that although relation (4.9) initially suggests a decreasing trend of spreading time with $Bo_b$, a closer inspection reveals that the capillary–inertial timescale in the numerator ($\tau_{CI} \propto R_b^{3/2}$) dominates over the $Bo_b$ dependent denominator ($Bo_b \propto R_b^2$), leading to an overall scaling of $t_{sp} \propto R_b^{1/2}$. This explains the increasing trends of $t_{enc}$ with $Bo_b$ in Figure 21 (b). Overall, the encapsulation time trends in Figures 20 and 21 confirm that the numerical results are consistent with the theoretical scaling laws governing spreading time and velocity.

## 5. Summary and conclusion

The dynamic interaction between three immiscible fluid phases remains one of the most intricate and compelling challenges in fluid mechanics. This complexity arises from the simultaneous action of capillary, viscous, inertial, and gravitational forces, operating across multiple spatial and temporal scales. This phenomenon has been explored in various contexts, such as bubbles interacting with pool-droplet interfaces (Li *et al.* 2014a, 2014b; Maës *et al.* 2025), aqueous droplet engulfment in lighter oils layer (Cuttle *et al.* 2021; Maës *et al.* 2025), and more recently, the behavior of laser-induced cavitation bubbles within droplets or hemispherical pendant drops (Li *et al.* 2024; Ren *et al.* 2023). These studies reveal rich, multi-scale interfacial dynamics influenced by both physicochemical and hydrodynamic factors. Here, we investigated a less explored yet equally intricate scenario: the engulfment of an air bubble by a rising oil droplet within a host liquid. By integrating lattice Boltzmann simulations with high-speed experiments and theoretical scaling, we systematically examined how various parameters, including droplet/medium viscosity, density ratio, surface tension, bubble-to-droplet size ratio, gravitational acceleration, and off-center impact, govern the bubble encapsulation dynamics. Using the phase-field index to track the spreading front of the droplet, we quantified the bubble encapsulation time ($t_{enc}$) and analyzed their velocity profiles and shape evolution throughout the process.

The encapsulation process was found to unfold in four primary stages: (1) impact and film drainage, driven by hydrodynamic and intermolecular forces; (2) droplet spreading and engulfment, where released interfacial energy sharply accelerates the bubble, especially at low viscosity; (3) bubble reshaping and ascent within the droplet, governed by surface tension resisting viscous, inertial, and gravity-induced deformation; and (4) compound ascent, where buoyancy balances viscous drag. A transitional saddle region appears between stages (2) and (3), particularly pronounced in low-viscosity, low $S_o$ values, or cases where the bubble undergoes notable shape deformations. As the spreading coefficient ($S_o$) approaches zero, the second and third stages overlap, with oscillatory velocity profiles indicating strong energy exchange between capillary and kinetic modes that lead to repeated shape deformations. Our parametric study reveals that higher $S_o$ reduces the encapsulation time ($t_{enc}$) by promoting faster capillary-driven spreading while increasing viscosity in either the drop or the medium exponentially prolongs encapsulation by suppressing interfacial motion. Shape analysis of the coated bubble shows that as the drop-to-medium viscosity increases, viscous stresses progressively dominate over capillary-inertial forces, resulting in a rounder compound morphology ($d_h/d_v \rightarrow 1$ at higher viscosities). Droplet density, however, exhibited a milder influence on the engulfment process, with $t_{enc}$ increasing as the drop density approached that of the medium, consistent with slower neck growth predicted by theory. Moreover, for equal-sized cases, smaller droplets were found to reduce $t_{enc}$ due to higher capillary pressure.

In the low viscosity regime ($Oh_s < 0.1$), both experiments and LBM simulations indicate that for the droplets equal to, or larger than the bubble, the neck radius grows with a power-law exponent of 0.44, confirming the proposed approximation in Li *et al.* (2014a). For the droplet smaller than the bubble, the exponent increases to 0.5, as reported for the coalescence of immiscible or homogeneous droplets (Xu *et al.* 2024). Variation in drop-to-bubble ratio ($R_d/R_b$) at intermediate Ohnesornge numbers ($Oh_s < 0.38$) demonstrates that although smaller droplets exhibit faster neck growth, $t_{enc}$ increases for $R_d/R_b < 1$ by enlarging bubble radii, reflecting the bigger surface that needs to be covered. For $R_d/R_b > 1$, $t_{enc}$ plateaus with increasing droplet size, closely resembling the case of a bubble passing through a liquid-liquid interface. Interestingly, size asymmetry also introduces a directional fluid flow: when $R_d/R_b < 1$, the droplet pulls the bubble during the spreading stage, driven by higher pressure within the smaller drop. Conversely, for bigger drops (e.g. $R_d/R_b = 2$), the bubble appears to be engulfed by the drop, with the pressure gradient driving flow from bubble to drop. Simulations indicate a regime shift between spherical and mildly deformed morphology around $Bo_b \approx 0.4$ and $Bo_d \approx 0.1$, where the horizontal-to-vertical aspect ratio ($d_h/d_v$) of the compound exceeds 1.1 under elevated $Bo$ and lower oil volume fractions. Examining off-center impacts, we detect an unexpected minimum in encapsulation time at an intermediate impact parameter ($B = 0.5$, or ~45° angle). This optimal configuration appears to result from enhanced momentum transfer between vertical and horizontal directions. At elevated gravity, the boundary between engulfment and rebound shifts toward lower impact parameters due to stronger bubble-induced shear forces deflecting the drop.

To consolidate our findings, we developed a theoretical scaling framework that incorporates the spreading coefficient $S_o$ in place of surface tension, reflecting its critical role in ternary fluid wetting. In the spherical regime ($Bo_{b,d} < 0.11$), droplet spreading speed was found to scale with a viscous-capillary velocity law $U_{VC} \propto S_o/\mu$, where $t_{enc}$ increased with the Ohnesorge number ($Oh_s$). In the deformed regime ($0.11 < Bo_b < 2.2$), $t_{enc}$ is correlated with a capillary-gravitational timescale ($\tau_{CG}$) dependent on both capillary-inertial time and $Bo$. Interestingly, a contrasting trend manifested in the relationship between $t_{enc}$ and $Bo$: increasing $Bo_b$ via gravity (at fixed size) tends to reduce $t_{enc}$ owing to the enhanced buoyancy-driven motion, while increasing $Bo_b$ by enlarging the bubble radius ($R_b$) prolongs it due to the greater interfacial area involved.

**Funding statement.** This research received no specific grant from any funding agency, commercial or not-for-profit sectors.

**Conflict of Interest.** The authors report no conflict of interest.

**Data Availability.** The data that support the findings of this study are available from the corresponding author, S.M. Hosseinalipour, upon reasonable request.

**Author ORCIDs.**
Adel Ebadi https://orcid.org/0000-0001-6106-2211;
Mostafa S. Hosseinalipour https://orcid.org/0000-0002-6128-8920.

**Author contributions.** S.M. H. funded and supervised the project; A.E. designed and performed the research; R.K. contributed to development of the numerical model; F.A. and H.A.A. contributed to the construction of the experimental set-up and performing the tests; A.E. wrote the manuscript with feedback from co-authors, which was revised and edited by S.M.H.